\documentclass{aa}  
\usepackage{txfonts}
\usepackage{graphicx,subcaption}
\usepackage{natbib,twoopt}
\usepackage{xcolor}
\usepackage{multirow} 

\usepackage[breaklinks=true]{hyperref} %% to avoid \citeads line fills
\hypersetup{
  colorlinks=true,   %% links colored instead of frames
  urlcolor=blue,     %% external hyperlinks
  linkcolor=blue,     %% internal latex links (eg Fig)
  citecolor=blue
}

\usepackage{orcidlink}
\newcommand{\msun}{\mbox{M$_{\odot}$}}
\newcommand{\rsun}{\mbox{R$_{\odot}$}}
\newcommand{\lsun}{\mbox{L$_{\odot}$}}

\newcommand{\teff}{\textit{T$_{\rm eff}$}}

\newcommand{\Mdot}{$\dot{M}$}

\providecommand{\sun}{_\text{\sun}}  

\makeatletter
\renewcommand*\aa@pageof{, page \thepage{} of \pageref*{LastPage}}
\makeatother

\begin{document}

   \title{Episodic mass loss in the very luminous red supergiant \\ $[$W60]~B90 in the Large Magellanic Cloud}
 %  \subtitle{I. Overview of observations and stellar content}

   \author{G. Munoz-Sanchez\inst{\ref{noa}, \ref{nkua}}\orcidlink{0000-0002-9179-6918}
        \and
        S. de Wit\inst{\ref{noa}, \ref{nkua}}\orcidlink{0000-0002-9818-4877} 
        \and
        A.Z. Bonanos\inst{\ref{noa}}\orcidlink{0000-0003-2851-1905}
        \and
        K. Antoniadis\inst{\ref{noa}, \ref{nkua}}\orcidlink{0000-0002-3454-7958} 
        \and
        K. Boutsia\inst{\ref{noirlab}, \ref{lco}} \orcidlink{0000-0003-4432-5037} 
        \and \\
        P. Boumis\inst{\ref{noa}}\orcidlink{0000-0002-4260-940X} 
        \and
        E. Christodoulou\inst{\ref{noa}, \ref{nkua}}\orcidlink{0000-0003-4332-3646} 
        \and
        M. Kalitsounaki\inst{\ref{noa},\ref{nkua}}
        \and
        A. Udalski\inst{\ref{warsaw}}
        }

   \institute{IAASARS, National Observatory of Athens, Metaxa \& Vas. Pavlou St., 15236, Penteli, Athens, Greece\label{noa}
        \and
        Department of Physics, National and Kapodistrian University of Athens, Panepistimiopolis, Zografos, 15784, Greece\label{nkua}
        \and
        %Institute of Astrophysics, FORTH, GR-71110, Heraklion, Greece\label{crete}
        %\and
     Mid-Scale Observatories/NSF NOIRLab, 950 N. Cherry Ave., Tucson, AZ 85719, USA\label{noirlab}
     \and
     Las Campanas Observatory, Carnegie Observatories, Colina El Pino, Casilla 601, La Serena, Chile\label{lco}
     \and
     Astronomical Observatory, University of Warsaw, Al. Ujazdowskie 4, 00-478 Warsaw, Poland\label{warsaw}
     }
   \date{}

% \abstract{}{}{}{}{} 
% 5 {} token are mandatory
 
  \abstract
 {Despite mounting evidence that extreme red supergiants (RSGs) undergo episodic mass-loss events, their role in RSG evolution remains uncertain. Critical questions remain unanswered, such as whether or not these events can strip the star, and their timescale and frequency.}
 {This study delves into [W60]~B90, one of the most luminous and extreme RSGs in the Large Magellanic Cloud (LMC), with our aim being to search for evidence of episodic mass loss. Our discovery of a bar-like nebular structure at 1~pc, which is reminiscent of the bar around Betelgeuse, raised the question of whether [W60]~B90 also has a bow shock, motivating the present study.}
 {We collected and analyzed proper motion data from \textit{Gaia}, as well as new multi-epoch spectroscopic and imaging data, and archival time-series photometry in the optical and mid-infrared (MIR). We used \textsc{marcs} models to derive the physical properties of the star from the spectra.}
{We find [W60]~B90 to be a walkaway star, with a supersonic peculiar velocity in the direction of the bar. We detect shocked emission between the bar and the star, based on the [S~\textsc{ii}]/H$\alpha$ > 0.4 criterion, providing strong evidence for a bow shock. The 30~yr optical light curve reveals semi-regular variability, showing three similar dimming events with $\Delta V\!\sim\! 1$~mag, a recurrence of $\sim$12~yr, and a rise time of 400~days. We find the MIR light curve to vary by 0.51~mag and 0.37~mag in the WISE1 and WISE2 bands, respectively, and by 0.42~mag and 0.25~mag during the last dimming event. During this event, optical spectroscopy reveals spectral variability (M3~I to M4~I), a correlation between the \teff\ and the brightness, increased extinction, and, after the minimum, spectral features incompatible with the models. We also find a difference of $>$300~K between the \teff\ measured from the TiO bands in the optical and the atomic lines from our $J$-band spectroscopy.} 
{[W60]~B90 is a more massive analog of Betelgeuse in the LMC and therefore the first single extragalactic RSG with a suspected bow shock. Its high luminosity of $\log(L/$\lsun$)=5.32$~dex, mass-loss rate, and MIR variability compared to other RSGs in the LMC indicate that it is in an unstable evolutionary state, undergoing episodes of mass loss. Investigating other luminous and extreme RSGs in low-metallicity environments using both archival photometry and spectroscopy is crucial to understanding the mechanism driving episodic mass loss in extreme RSGs in light of the Humphreys-Davidson limit and the ``RSG problem''.}

   \keywords{Stars: individual: [W60] B90, stars: massive - stars: supergiants - stars: atmospheres - stars: late-type - stars: mass-loss}

   \titlerunning{The very luminous red supergiant [W60] B90}
   \authorrunning{Munoz-Sanchez et al.}

   \maketitle
%
%-------------------------------------------------------------------

\section{Introduction}\label{sec:intro}

Red supergiants (RSGs) are evolved massive stars with initial masses of $8-25$~\msun~\citep{Ekstrom2012, Levesque2017}. During their evolution, RSGs increase their luminosity and therefore manifest larger radii and cooler temperatures, before ending their life by exploding as a supernova (SN) or collapsing directly into a black hole \citep{Smartt2015, Sukhbold2016, Adams2017, Laplace2020}. RSGs are more prone to exhibiting spectral-type variability as they become more luminous and cooler as a consequence of a more unstable state \citep[e.g.,][]{Levesque2007, Dorda2021}. Despite the fact that the RSG phase represents only 10\% of the lifetime of these stars, most stellar mass loss takes place during this phase of the stellar evolution. Empirical relations have revealed a robust correlation between luminosity and mass loss within the RSG phase \citep[e.g.,][]{deJager1988, vanLoon2005, Beasor2020, Humphreys2020,Yang2023, Antoniadis2024}. Evidence of episodic mass ejections has been found around luminous RSGs \citep[e.g., NML Cyg, VY CMa, Betelgeuse, RW Cep;][]{Richards1996, Humphreys2005, Humphreys2007, Decin2006, Dupree2022, Anugu2023}. These ejections are associated with gaseous outflows related to surface activity \citep{Humphreys2022}, which impact the photometric variability in terms of dimming events \citep[e.g.,][]{Guinan2019, Humphreys2021, Anugu2023}. Especially remarkable was the "Great Dimming" of Betelgeuse when the star unexpectedly decreased its brightness by one magnitude. A multiwavelength follow-up found this phenomenon to be the result of a mass-ejection event that formed dust and obscured the star \citep{Dupree2020, Montarges2021, Drevon2024}. \citet{Montarges2021} determined the ejected mass to be between 3 and 120$\%$ of the annual mass lost by Betelgeuse, demonstrating that the significance of episodic mass loss is uncertain by two orders of magnitude.\citet{Humphreys2022} revealed that the episodic outflows of Betelgeuse contribute significantly to its overall mass-loss history. On the other hand, in more extreme RSGs, such as VY CMa, the episodic ejections alone explain the high average mass-loss rate measured for the star.

Despite being one of the brightest stars in the sky, Betelgeuse exhibits many properties that remain unexplained \citep[see e.g.,][]{Wheeler2023}. The bow shock and bar structure in the vicinity of the star are particularly unusual \citep{Noriega-Crespo1997}. The origin of the bar is uncertain: some arguments support the relation to Betelgeuse \citep{Mackey2012} while others advocate for an interstellar origin \citep{Decin2012, Meyer2021}. On the contrary, the physics behind the bow shock is well understood, as it is produced when a star moves supersonically and the stellar wind interacts with the interstellar medium (ISM), creating an arc-like shape. Although they are commonly seen in OB runaways \citep{Noriega-Crespo1997OB}, only two other cases of Galactic single RSGs are known: $\rm{\mu}$~Cep and IRC-10414 \citep{Cox2012, Gvaramadze2014}. Their detection is useful for constraining the properties of the local ISM and the stellar wind \citep{Hollis1992, Kaper1997}. Because of the lack of a sophisticated grid of RSG models that include the wind, deriving the mass-loss rate, \Mdot, directly from optical spectroscopy is still impossible. Instead, the \Mdot~of RSGs is commonly derived from the mid-infrared (MIR) excess of the spectral energy distribution (SED) \citep[e.g.,][]{Riebel2012, Yang2023, Antoniadis2024}. However, this methodology depends on general assumptions such as the gas-to-dust ratio, dust grain size, and the mechanism of the wind, which introduce large uncertainties in the results. Therefore, the bow shock in RSGs provides a unique and independent scheme to estimate the \Mdot~and compare it with that obtained using other methods.   

[W60]~B90 is one of the most luminous RSGs in the Large Magellanic Cloud (LMC). It was reported by \citet{deWit2023}, within the ASSESS project \citep[Episodic mAss loSS in Evolved maSsive Stars]{Bonanos2024}, to have extreme parameters similar to those of WOH~G64 \citep{Levesque2009}. Our discovery of a bar-like structure, similar to the bar of Betelgeuse, at 1~pc from the star in an archival \textit{Hubble} Space Telescope (HST) image, immediately raised the question of whether [W60]~B90 could be the first extragalactic RSG with a bow shock. Moreover, its high luminosity close to the observed upper limit of RSGs in the LMC \citep[$\log(L/\lsun)=5.50$~dex;][]{Davies2018} and its high mass-loss rate within the LMC \citep[\Mdot~$=5.07\times10^{-6}$~\msun~yr$^{-1}$,][]{Antoniadis2024} indicate that this RSG is in an evolved evolutionary state and is undergoing considerable mass loss.

In this paper, we present a detailed analysis of [W60]~B90. We collected archival photometry to study the long-term photometric variability and we performed a multi-epoch spectroscopic campaign both to study its current spectral variability and search for evidence of shocked material in the circumstellar environment.
In Sect.~\ref{sec:observations}, we describe the observations obtained and the archival data used. In Sect.~\ref{sec:evidence_shocked_material}, we investigate the membership of [W60]~B90 to the LMC and its relation to the bar. We analyze the spectroscopic data for the circumstellar nebular emission and examine its shocked origin. In Sect.~\ref{sec:gen_variability}, we analyze the light curve, the variability in the optical and the MIR, and present the results derived from our optical and near-infrared (NIR) spectroscopic observations. We discuss the results and evolutionary status of the star in Sect.~\ref{sec:discusion_gen}, and summarize our conclusions in Sect.~\ref{sec:conclusions}.

%------------------------------------------------------------------
%------------------------------------------------------------------
%------------------------------------------------------------------

\section{Observations and data reduction}\label{sec:observations}

We discovered a nebular bar structure located at 4$\arcsec$ from [W60]~B90 on an HST image available in the Hubble Legacy Archive\footnote{\url{https://hla.stsci.edu/}} (see Fig.~\ref{fig:PM_LMC3}). The observations were obtained on 2007 July 27 UT05:34:16 under the program ID 10583, with an exposure time of 1000s and the F675W filter. Below, we describe the archival photometry collected and our spectroscopic observations of [W60]~B90.  

\subsection{Photometric catalogs}\label{sec:phot_catalogs}

We assembled the light curve of [W60]~B90 by compiling archival photometry\footnote{Updated up to the submission date of the paper.}  spanning $\sim$30~yrs from the following surveys: ATLAS forced photometry \citep{ATLASmain2018, ATLASvariable2018, ATLASserver2021}, ASAS \citep{ASAS1997}, ASAS-SN \citep{ASASSN2014_1, ASASSN2017_2}, \textit{Gaia} DR3 \citep{Gaiamision2016, GaiaDR32023}, the MACHO project \citep{MACHO1997}, NEOWISE \citep{NEOWISE2011}, OGLE \citep{Udalski1997, OGLEIII2008, Udalski2015}, the \textit{Spitzer} SAGE and SAGE-var projects \citep{SAGE2006, SAGEvar2015}, and \textit{WISE} \citep{WISE2010, ALLWISE2014}. We applied the following criteria to each survey to select the most reliable data:  

\begin{itemize}
    \item ATLAS: We used forced photometry on reduced images to obtain the light curve from the server. We used data points with flag $err=0$, $chi/N<100$, and an error below 0.1~mag.  
    
    \item ASAS: We used the photometry based on the smallest aperture (2~px) to avoid contamination from other sources due to the small pixel scale of the instrument (16$\arcsec$px$^{-1}$). We used the mean data, (i.e., B flag-category) to obtain a cleaner light curve. 
    
    \item ASAS-SN: We retrieved the data from Sky Patrol, selecting the image subtraction with the reference flux added option as it uses co-added data, considerably decreasing the photometry error. We rejected epoch photometry with errors above 0.1~mag, and removed the \textit{bm} camera data in the $g$-band, due to a systematic offset in the flux with respect to the other cameras.
    
    \item MACHO: We only used the $V_{KC}$-band as the star saturates in the $R_{KC}$-band. The MACHO calibration uses the $V_{KC}-R_{KC}$ color \citep[see Eq. 1 and 2 in][]{MACHOcalibration1999}, and therefore we computed the $V_{KC}$-band considering the approximation $V_{KC} \approx V_{M,t}+a_0+2.5\log(ET)$, where $a_0$ is the zero-point coefficient and $ET$ is the exposure time. We applied a $3\sigma$ clipping to discard outliers with small errors. 

    \item NEOWISE: [W60]~B90 has been observed since the mission started in 2014 during 20 epochs, each lasting over one week. We binned all the photometric measurements within an epoch by taking the median value and the uncertainty of the median. We discarded data with \textit{qual\_frame}~=~0 and \textit{chi}$^2>20$. However, NEOWISE photometry differs from ALLWISE for targets brighter than $W1<8$~mag and $W2<7$~mag. Hence, we applied an offset according to Fig. 6 in \citet{NEOWISEoffset2014} to correct the magnitudes.

    \item OGLE: We used $I$-band data from the OGLE-III shallow survey \citep{Ulaczyk2013}, and $V$-band data from the OGLE-II, OGLE-III, and OGLE-IV databases. Unfortunately, the star is saturated in the images of the main $I$-band monitoring of the LMC by OGLE, conducted continuously over the last 27 years.

    \item \textit{Spitzer}: We collected the epoch photometry from the SAGE and SAGE-var projects. 

\end{itemize}

\subsection{Optical spectroscopy}\label{sec:longslit_obs}

We performed spectroscopic observations with the Magellan Echellette (MagE) spectrograph \citep{MagE2008} placed on the 6.5~m Baade telescope at Las Campanas Observatory, Chile. We used the $0.7\arcsec\times10\arcsec$ long-slit, providing a wavelength coverage of $3500-9500$\r{A}, a spectral resolution of $R\sim5000$, and a spatial resolution of $0.3\arcsec$~px$^{-1}$ with binning $1\times1$. \autoref{tab:slit_obs} shows the coordinates of the slit center, UT date of the observations, the instrument, the exposure time, the position angle (PA) of the slits, and the slit width. The slits labeled Epoch1-Epoch4 were centered on the RSG, while slits Neb1-Neb6 were placed around the star to search for shocked material. We used the MagE pipeline \citep{Kelson2000, Kelson2003} for the bias and flat correction. We continued the reduction based on the \textsc{echelle} package of IRAF\footnote{\textit{IRAF} is distributed by the National Optical Astronomy Observatory, operated by the Association of Universities for Research in Astronomy (AURA) under agreement with the National Science Foundation.}, as we detected small artifacts in later steps of the MagE pipeline that could compromise the nebular emission. Finally, we used a flux standard to calibrate in flux the spectra with the IRAF routines \textsc{standard} and \textsc{sensfunc}.

We divided each slit into small sections to investigate the spatial distribution of the nebular emission. We extracted several apertures of 3~px ($0.9\arcsec$ or $\sim$0.25~pc for the distance of the LMC) in areas of the slits without contamination from background sources (see Table~\ref{sec:apendix_aper_flux}.1). We consider the 3~px aperture optimal as it minimizes the stellar contamination, achieves a better spatial resolution of the analysis of the circumstellar material (CSM), provides an adequate signal-to-noise ratio (S/N), and avoids a large impact from artifacts. The slits labeled Epoch1 and Neb1-3 did not have acquisition images. We obtained their position and orientation using the header parameters instead. 

%PA in the table is ROT-44.5 in the header of each image
\begin{table*}
\centering
\caption{Log of the long-slit observations with the 6.5~m Baade telescope}\label{tab:slit_obs}
\small
\begin{tabular}{l c c c c c r c}
\hline\hline
Slit name		& RA		& Dec.	 	& UT Date & Instrument & Exp. time	& PA$^a$  & Slit width 	\\
		& (J2000)	& (J2000)	& & & (s) & (\textdegree) &  ($\arcsec$)\\
\hline \\[-9pt]
Epoch1$^{b}$	& 05:24:19.31   & $-$69:38:49.4	& 2020 March 08 & MagE & $3\times180 + 1\times240$ & 60 & 1.0\\
EpochJ	& 05:24:19.31 & $-$69:38:49.3 & 2021 January 30 & FIRE &  $4\times15$ & 13 & 0.6\\
Epoch2	& 05:24:19.24   & $-$69:38:50.2	& 2022 December 02 & MagE & $3\times180 + 1\times240$ & 17 & 0.7\\
Epoch3   & 05:24:19.29   & $-$69:38:49.4  & 2023 April 07 & MagE & $3\times180 + 1\times240$ & 90 & 0.7 \\
Epoch4   & 05:24:19.29  & $-$69:38:49.7	& 2023 September 28 & MagE & $3\times180$ & 90 & 0.7 \\ \\[-9pt]
\hline \\[-8pt]
Neb1 & 05:24:20.22   & $-$69:38:48.2 & 2022 March 20 & MagE & $3\times400$ & 77 & 0.7\\
Neb2 & 05:24:20.06   & $-$69:38:48.1 & 2022 March 20 & MagE & $3\times400$ & 120 & 0.7\\ 
Neb3 & 05:24:20.00   & $-$69:38:45.4 & 2022 March 20 & MagE & $3\times400$ & 79 & 0.7\\
Neb4 & 05:24:19.60   & $-$69:38:47.0 & 2022 December 02 & MagE & $3\times400$ & 111 & 0.7\\
Neb5 & 05:24:20.00   & $-$69:38:51.8 & 2022 December 02 & MagE & $3\times400$ & 9	& 0.7\\
Neb6 & 05:24:19.94  & $-$69:38:45.0 & 2023 September 28 & MagE & $3\times400$ & 160 & 0.7\\

\hline
\end{tabular}
\tablefoot{
\tablefoottext{a}{The position angle PA $=0$\textdegree~refers to the South-to-North direction, while PA $=90$\textdegree~is the West-to-East direction.}\\
\tablefoottext{b}{From \citet{deWit2023}.}}
%\tablebib{}
\end{table*}

\subsection{Near-infrared spectroscopy}\label{sec:nearIR_obs}

We observed [W60]~B90 with the Folded-port InfraRed Echellete (FIRE) instrument placed on the 6.5~m Baade telescope at Las Campanas Observatory, Chile, on 2021 January 30 (Table~\ref{tab:slit_obs}). Using the $0.6\arcsec$ slit and the binning 1$\times$1, we covered the range $0.84-2.4$~$\mu$m with a spectral resolution of $R\sim6000$ and pixel scale of $0.15\arcsec$~px$^{-1}$. We obtained four exposures of 15 seconds with the high-gain mode following an ABBA pattern. We reduced the data with the official FIRE pipeline developed in IDL\footnote{\url{https://wikis.mit.edu/confluence/display/FIRE/FIRE+Data+Reduction}}, including the telluric correction and the flux calibration of the spectra. 

\subsection{Spitzer spectroscopy }\label{sec:data_spit}

[W60]~B90 was observed spectroscopically by \textit{Spitzer} with the Infrared Spectrograph (IRS) \citep{Houck_2004} inside the programs ID1094 (AORKEY: 6076928) and ID40061 (AORKEY: 22272512). Four modules were used (Short-Low, Short-High, Long-Low, and Long-High) covering the spectral region from 5.2~$\mu$m to 38~$\mu$m and providing low ($R\sim60-130$) and high ($R\sim600$) resolution for the short and long configurations.  However, we used only the Short-Low of Program ID1094 from Level 2 of the Post Basic Calibrated Data (PBCD) because an unidentified nearby source contaminated the longer wavelengths. This source is comparable in brightness to [W60]~B90 only in MIPS1 (24~$\mu$m) images. The data from program ID40061 were discarded, as the star was not properly centered on the slit on the short-low exposures. Finally, we added the synthetic photometry at 12 and 16 $\mu$m provided in the IRS table to include them in the SED fitting (see sect.~\ref{sec:dis_mdot}).

%------------------------------------------------------------------
\section{Evidence of shocked material} \label{sec:evidence_shocked_material}  

\subsection{LMC membership and proper motion} \label{sec:proper_motion}

\textit{Gaia}~DR3 reported a parallax $0.0457\pm0.0270$~mas for [W60]~B90, which corresponds to a distance of $22\substack{+32 \\ -8}$~kpc. This uncertainty prevented us from determining whether [W60]~B90 belongs to the LMC or our Galaxy. \citet{deWit2023} previously analyzed the membership to the LMC based on the radial velocity (RV) from Ca~\textsc{ii} triplet, its position within the LMC and \textit{Gaia}~DR2. We go one step further and use the kinematic analysis of the LMC by \citet{Jimenez2023}, based on \textit{Gaia}~DR3. These authors derived a probability $P_{LMC}$ for stars within the field of the LMC to belong to that galaxy, and report $P_{LMC}=0.87$ for [W60]~B90, with $P_{LMC}=0$ corresponding to a foreground star, $P_{LMC}=1$ to a highly probable LMC member, and $P_{LMC}=0.52$ as the class cut-off limit. Considering also the RV of 263~km~s$^{-1}$ reported by \textit{Gaia}~DR3, we conclude that [W60]~B90 is a genuine LMC member. 

Next, we investigated whether or not [W60]~B90 and the bar are physically related. We computed the peculiar velocity of the RSG, to determine whether it moves in the direction of the bar. The motion towards the bar would be consistent with a bow shock scenario, where the interaction of the wind with the ISM causes shock-ionization. We used \textit{Gaia} DR3 to collect the proper motions (PMs) of all the stars within a cone radius of $1.8\arcmin$ to $36\arcmin$ centered W60 B90. We then cleaned the sample from foreground contamination using \citet{Jimenez2023}, and obtained a median value of the PM within the cone. This value was then subtracted from the PM of [W60]~B90 to obtain the local motion. Finally, we tested the robustness of our analysis by exploring different cone sizes and probability thresholds to clean the foreground contamination. We selected $36\arcmin$, $18\arcmin$, $6\arcmin$, $4.5\arcmin$, $3\arcmin$, $2.4\arcmin$, and $1.8\arcmin$ for the cone sizes, which correspond to local distances of 520, 260, 87, 65, 43, 35, and 26~pc, assuming a distance $D=49.59$~kpc to the LMC \citep{Pietrzynski2019}. We considered the probability thresholds $P_{LMC}$ = 0.5, 0.7, and 0.9. Table~\ref{sec:apendix_PM_values}.1 lists the derived PM and peculiar velocity of [W60]~B90 as a function of the cone size and $P_{LMC}$. $N_{Gaia}$ is the number of stars inside the cone, while $N_{clean}$ is the number of stars remaining after the foreground cleaning. 

\begin{figure}
        \centering
   	\resizebox{0.97\hsize}{!}{\includegraphics{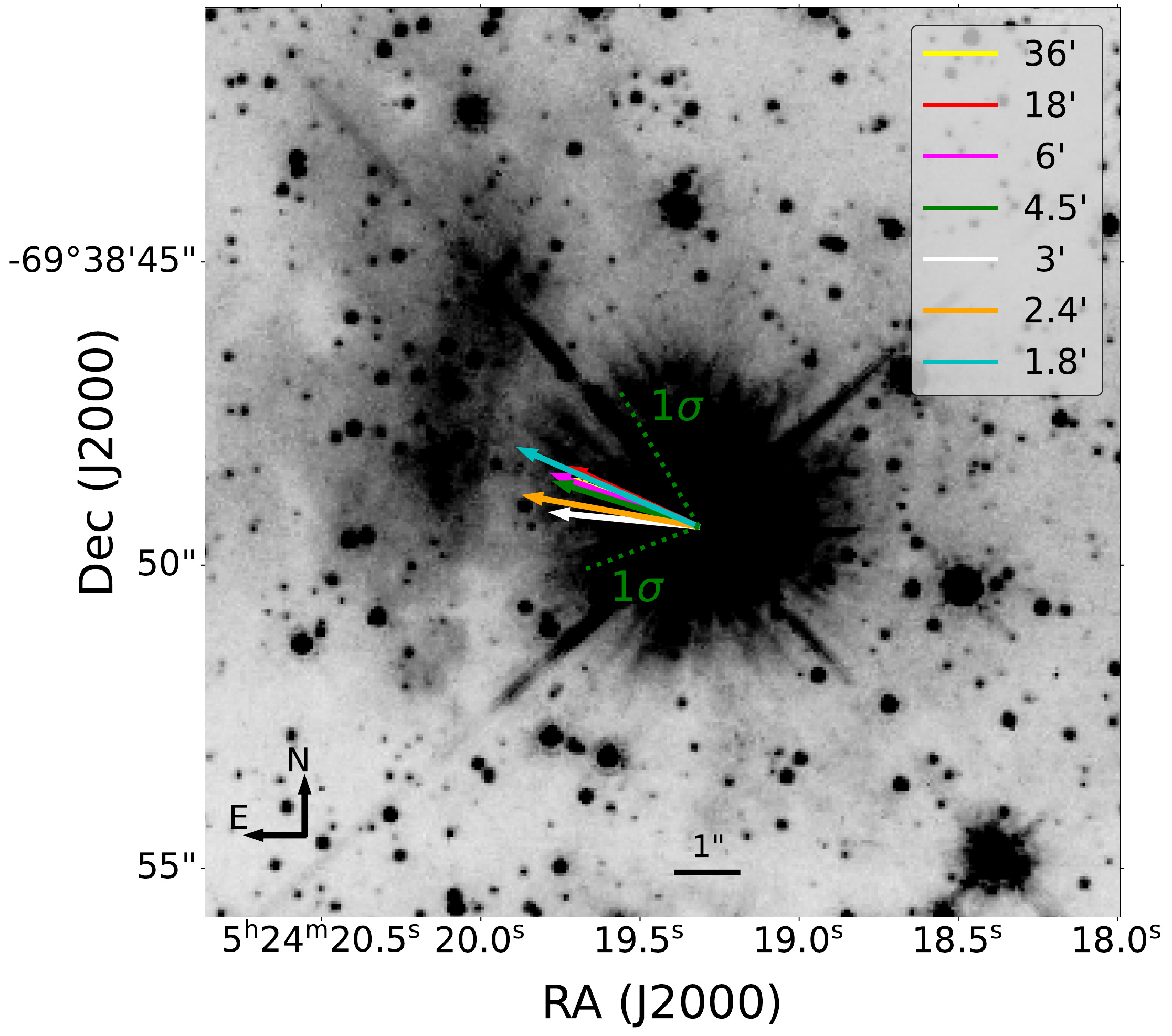}}
  	\caption{HST F675W image of [W60]~B90 showing the PM analysis for $P_{LMC}=0.7$. The arrows represent the peculiar velocity direction of [W60]~B90 for each cone to compute the local PM of the LMC. The length of the arrows is scaled with the peculiar velocity. The green dotted lines show the 1$\sigma$ PM error from \textit{Gaia} DR3 on the $4.5\arcmin$ cone size.}\label{fig:PM_LMC3}
\end{figure}

We present the results for $P_{LMC}=0.7$ in Fig.~\ref{fig:PM_LMC3} as an example. Our analysis reveals that the selected $P_{LMC}$ thresholds barely affect the results, while the direction of the peculiar velocity varies slightly, depending on the cone size. These variations are considerably smaller than the 1$\sigma$ error from \textit{Gaia} DR3, and all are consistent with the projected motion of the star towards the bar. We derived a peculiar velocity ranging from $16-25~(\pm$11)~km~s$^{-1}$ depending on the parameters assumed, but still compatible with moving faster than the speed of sound (see Sect.~\ref{sec:discusion_bow_shock}). Forthcoming \textit{Gaia} releases are needed to improve the accuracy of the PM, reducing the uncertainties on the orientation and absolute value of the peculiar velocity of [W60]~B90.

\subsection{The [S~\textsc{ii}]/H\texorpdfstring{$\alpha$}{a} ratio} \label{sec:SIIvsHa}

\begin{figure*}
    \centering
    \includegraphics[width=0.82\textwidth]{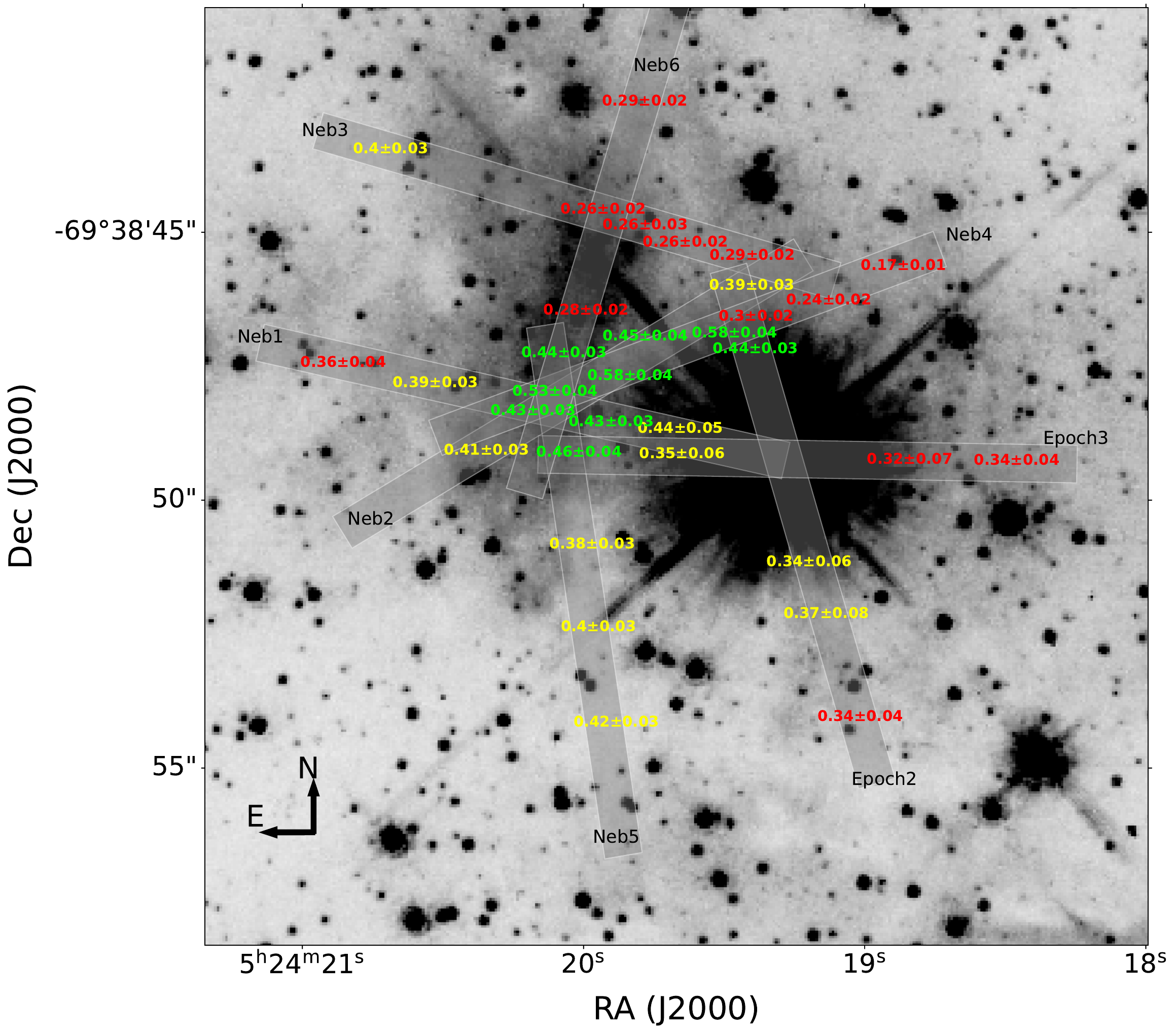}
    \caption{HST F675W image with the slits and the [S~\textsc{ii}]/H$\alpha$ ratio for each aperture overlaid. The green color corresponds to ratios above 0.4 within the error, the yellow color corresponds to ratios with a lower limit below 0.4, and the red color corresponds to apertures below 0.4 within the error.}\label{fig:SIIvsHa_fig}
\end{figure*}

Nebular emission can arise from the energy released in shocks or photoionizing radiation depending on the physical conditions. Historically, the ratio [S~\textsc{ii}]/H$\alpha$ has been used to separate between the two mechanisms, being photoionized when [S~\textsc{ii}]/H$\alpha<0.4$ and shocked when [S~\textsc{ii}]/H$\alpha\geq 0.4$ \citep{Mathewson1973A}. In H~\textsc{ii} regions, the radiation from hot stars ionizes sulfur mainly to S~\textsc{iii}. However, when shocks cause nebular emission, the energy is insufficient to produce S~\textsc{iii}, and S~\textsc{ii} becomes the main state of sulfur, increasing the [S~\textsc{ii}]/H$\alpha$ ratio. Although some works have identified shocked material with ratios down to [S~\textsc{ii}]/H$\alpha$~=~0.3 \citep[Fig. 5 in][]{Kopsacheili2020, Gvaramadze2014}, we considered 0.4 to be a more conservative criterion to confirm the presence of shocked material. 

We used the spectra presented in Table~\ref{tab:slit_obs} to analyze the nebular emission of the CSM. We measured the flux and the error of the emission lines with the IRAF task \textsc{splot}. We dereddened the fluxes using the Balmer decrement, which is the difference between the observed ratio of the intensities H$\alpha$/H$\beta$ with the theoretical ratio of 2.86  \citep[assuming $T_e=10^4$~K and $n_e=100$~cm$^{-3}$;][]{Osterbrock1989}. We used the Python tool \texttt{PyNeb} \citep{Luridiana2015} to calculate $E(B-V)$ and $c({\rm H}\beta)$ for each spectrum and dereddened them accordingly using the \citet{Gordon2003} extinction law for the LMC. We established a detection limit of $3\sigma$ to include a line in the analysis.  We report the locations of each aperture extracted in Table~\ref{sec:apendix_aper_flux}.1 and the measurements of the nebular emission in Table~\ref{sec:apendix_aper_flux}.2. In the latter table, we show the identified lines, their central wavelength, the observed and dereddened fluxes relative to H$\beta$, the S/N of the line, the RV, the extinction coefficients $E(B-V)$ and $c({\rm H}\beta)$, and the ratios of the ions.

We present the measured [S~\textsc{ii}]/H$\alpha$ ratios in the circumstellar environment of [W60]~B90 in Fig.~\ref{fig:SIIvsHa_fig}. We report values [S~\textsc{ii}]/H$\alpha>$~0.4 with 1$\sigma$ confidence (highlighted in green), which are in agreement with the PM direction and mainly concentrated between the bar and the star (see Sect.~\ref{sec:proper_motion}). The presence of the newly discovered B1V star at 7$\arcsec$ (see Appendix \ref{sec:apendix_Bstar}) and a nearby H~\textsc{ii} region at 14$\arcsec$ \citep[LHA 120-N 132B;][]{Henize1956} could explain the photoionization of the area and the nebular emission. However, if these stars were responsible for the nebular emission, the ratios should be homogeneous around our RSG and lower than 0.4-0.5. Moreover, the enhanced values in positions aligned with the PM of our RSG support the shocked mechanism from the interaction of [W60]~B90 with the CSM as the cause of the high [S~\textsc{ii}]/H$\alpha$ ratios. We also examine the inhomogeneity of the CSM around the RSG by extracting symmetric apertures at the North, South, East, and West positions. We used the Epoch2 and Epoch3 slits, which were centered on the star, and we took the peak of the star emission as a reference. We chose a distance of 7~px ($\sim2.1\arcsec$) from the peak as it was a good compromise between being close to the star and not having the nebular emission embedded in the continuum of the RSG. We compare the four positions in the lower left panel of  Fig.~\ref{fig:examples_SIIvsHa} to highlight the inhomogeneous emission at a distance of $2.1\arcsec$ ($\sim0.5$~pc) from [W60]~B90. We also show the comparison of the spectrum with the highest and lowest [S~\textsc{ii}]/H$\alpha$ ratio in the right panel of Fig.~\ref{fig:examples_SIIvsHa}.

\subsection{Other line ratios} \label{sec:other_ratios}

We used other line ratios to verify the origin of the nebular emission around [W60]~B90, including the ratios [O~\textsc{i}]/H$\alpha$, [O~\textsc{ii}]/H$\beta$, [O~\textsc{iii}]/H$\beta$, [N~\textsc{ii}]/H$\alpha$, and [S~\textsc{ii}]/H$\alpha$. In Fig.~\ref{fig:Kopsacheili_ratios}, we compare our measurements with the diagnostics presented in \citet{Kopsacheili2020}. The diagnostics for all the apertures agree with a shocked scenario, except for [O~\textsc{iii}], which indicates a photoionized mechanism. This may be explained by the fact that the models in \citet{Kopsacheili2020} only apply to shocks with velocities between 100--1000~km~s$^{-1}$, while the peculiar velocity of [W60]~B90 is $\leq 35$~km~s$^{-1}$. Material ejected from the star at such a low velocity will not contain enough energy to ionize O~\textsc{ii} to O~\textsc{iii}. 
The presence of a nearby H~\textsc{ii} region at 14$\arcsec$ \citep[LHA 120-N 132B;][]{Henize1956} and the newly discovered B1V star at 7$\arcsec$ (see Appendix \ref{sec:apendix_Bstar}) could also explain the photoionized nature of [O~\textsc{iii}]. Radiation from nearby hot stars might contribute to the ionization of the gas around our RSG and cause the emission in areas with low [S~\textsc{ii}]/H$\alpha$. Even in the shocked areas, nebular emission might arise from a combination of shocks and photoionization from nearby hot stars. Another explanation for the inconsistency of [O~\textsc{iii}] is the brightness of the line. In all the apertures,  the strength of [O~\textsc{iii}] $\lambda$5007 was approximately at the noise level ($3\sigma$ detection), and therefore the fluxes might be underestimated, leading to lower ratios. 

\subsection{Radial velocity of the CSM}

We measured the RV of the CSM from the central wavelengths of the Balmer lines and forbidden emission lines detected (see Table~\ref{sec:apendix_aper_flux}.2). We compared the RV of each ion at every location observed around [W60]~B90. However, the difference in the RV among the apertures was smaller than the $\pm7$~km~s$^{-1}$ error derived from the spectral resolution, preventing us from studying how the RVs are spatially distributed. We have, however, computed the median RV of each ion, using the individual velocities from each aperture (Table~\ref{tab:rad_vel_ion}). We grouped the lines [S~\textsc{ii}] \rm{$\lambda\lambda$}6717, 6731 and [O~\textsc{ii}] \rm{$\lambda\lambda$}3726, 3729 to compute a single RV for [S~\textsc{ii}] and [O~\textsc{ii}], respectively. Given that the RV of [W60]~B90 is $263.49\pm1.02$~km~s$^{-1}$ from \textit{Gaia}~DR3, we compared the kinematics of the nebular material to the kinematics of the star. All nebular lines are redshifted by $\sim$10~km~s$^{-1}$ compared to the star, except for [S~\textsc{ii}] and [O~\textsc{i}], which are $\sim$20~km~s$^{-1}$ and $\sim$40~km~s$^{-1}$, respectively. Each estimate is consistent with the low-velocity scenario ($\leq 100$~km~s$^{-1}$).

\begin{table}[!htb]
\caption{Mean radial velocity of the CSM} \label{tab:rad_vel_ion} 
\small
\renewcommand{\arraystretch}{1.4}

\centering
\begin{tabular*}{0.6\columnwidth}{cc|cc}
  \hline \hline
  Ion & RV & Ion & RV \\
   & (km s$^{-1}$) &  & (km s$^{-1}$) \\
  \hline
  H$\alpha$ & $275\pm2$ & [O~\sc{i}] & $310\pm4$ \\
  H$\beta$ & $276\pm3$ & [O~\sc{ii}] & $272\pm4$ \\
   $[$S~\sc{ii}] & $284\pm3$ & [O~\sc{iii}] & $276\pm5$ \\
   $[$N~\sc{ii}] & $277\pm4$ & & \\ 
  \hline
  \multicolumn{2}{r}{[W60] B90} & \multicolumn{2}{l}{$263.49\pm1.02$} \\
  \hline
\end{tabular*}
\centering
\tablefoot{Spectral resolution error is $\pm7$ km s$^{-1}$.}\\
\end{table}

%------------------------------------------------------------------
%------------------------------------------------------------------
%------------------------------------------------------------------
\section{Variability of [W60]~B90} \label{sec:gen_variability}

\subsection{Optical light curve} \label{sec:opt_lightcurve}

\begin{figure*}
   	\resizebox{1.0\hsize}{!}{\includegraphics{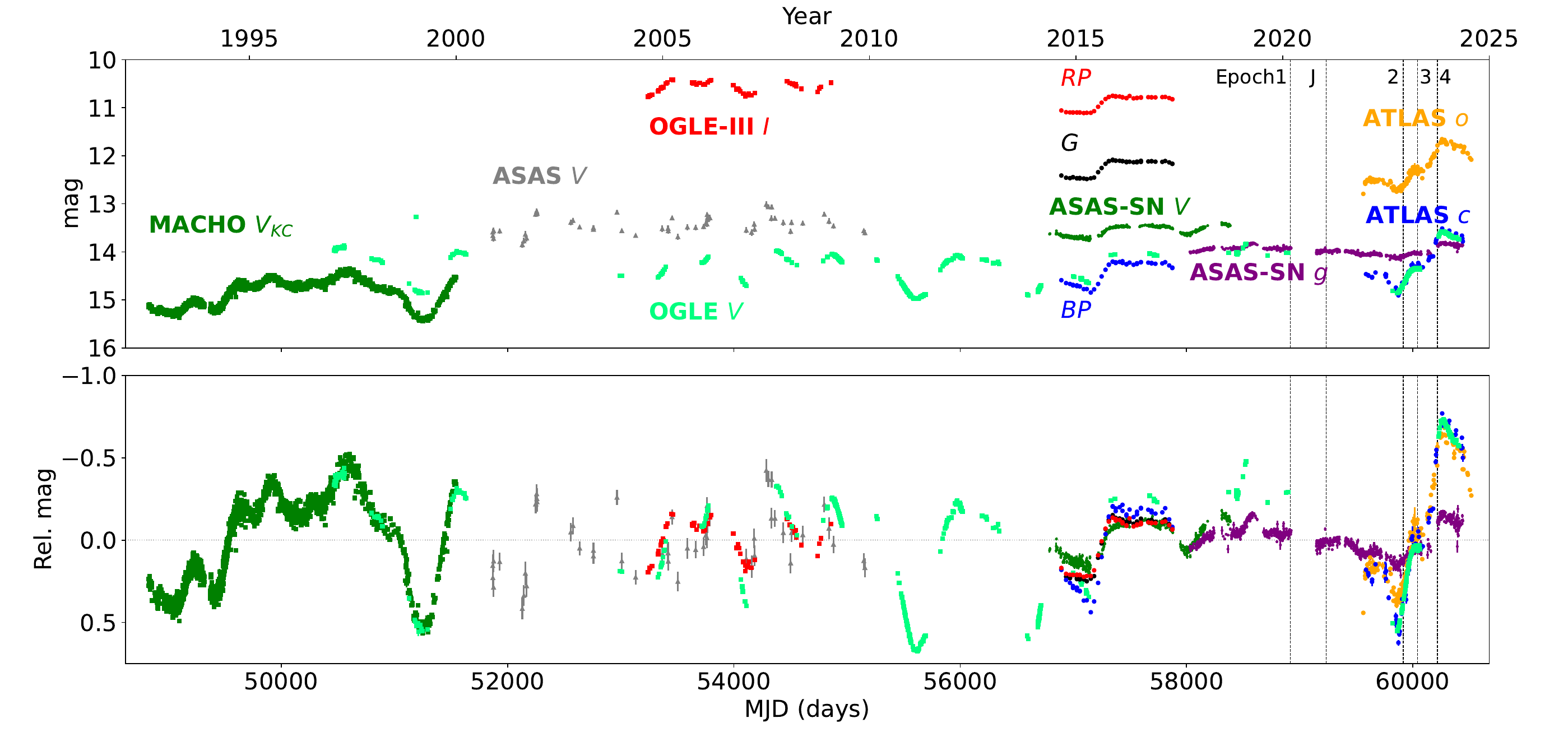}}
  	\caption{Light curve of [W60]~B90 indicating the respective survey and filter. \textit{Top}: Magnitudes obtained from each survey. \textit{Bottom}: Relative magnitude after subtracting the mean value of each data set. The vertical lines represent the epoch-spectroscopy as labeled in Table~\ref{tab:slit_obs}.}\label{fig:fig_lightcurve}
\end{figure*}

\begin{figure*}
        \centering
        \resizebox{1.0\hsize}{!}{\includegraphics{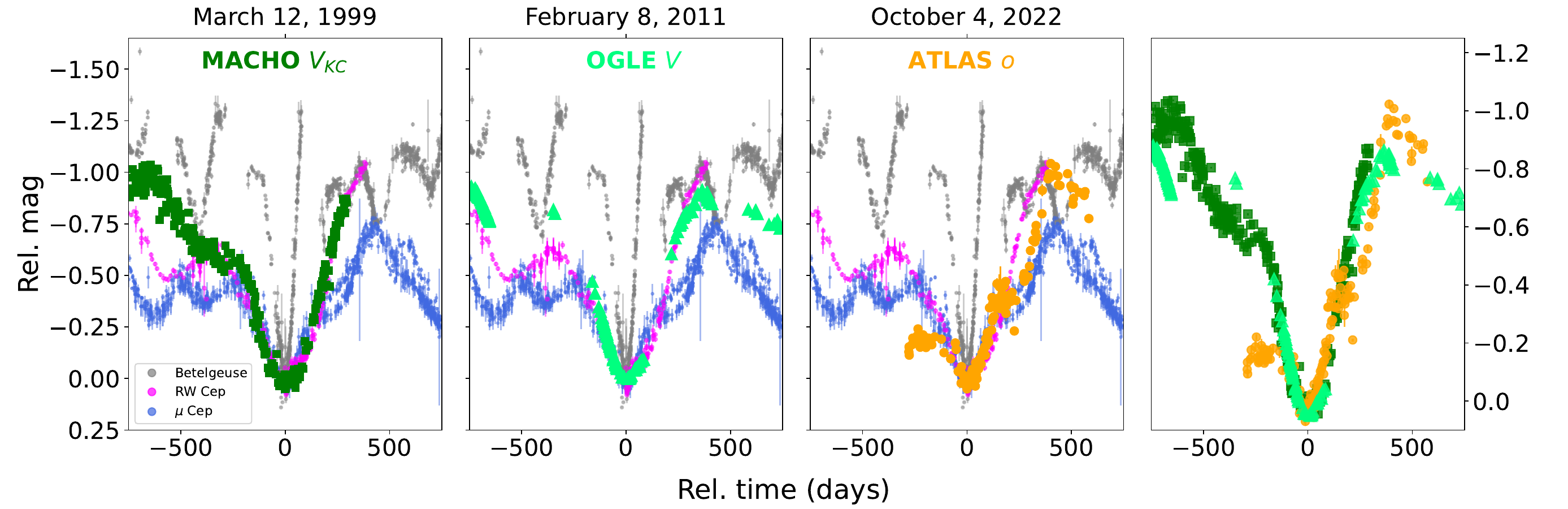}}
  	\caption{Zoom in on the dimming events of [W60]~B90. The first three panels show the comparison of the three dimmings of [W60]~B90 with the Great Dimming of Betelgeuse (gray), the dimming of RW Cep (magenta), and a minimum from $\mu$ Cep (blue) in the $V$-band from AAVSO. The right panel overplots the photometry of the three dimmings of [W60]~B90. We used the minimum of each data set as the zero-point reference for each graph.}\label{fig:fig_dimmings}
\end{figure*}

We assembled multi-epoch photometry of [W60]~B90 from 1992 to the submission date of this work and we present the optical light curve in Fig. \ref{fig:fig_lightcurve}. We subtracted the mean value of each data set to obtain a relative magnitude, allowing for a comparison between different filters and surveys (see lower panel of Fig.~\ref{fig:fig_lightcurve}). The amplitude of the ASAS-SN data is smaller than other surveys due to its pixel scale (8$\arcsec$px$^{-1}$). Nearby sources are blended with [W60]~B90, contaminating the data by adding a constant flux, which yields a lower amplitude. Nevertheless, we considered ASAS-SN as a guideline for the general shape of the light curve between 2017 and 2022, due to the lack of other data covering these years. After 2022, ATLAS and OGLE $V$-band provide better photometry due to their improved pixel scale (1.86 and 0.26$\arcsec$px$^{-1}$, respectively).

We classify [W60]~B90 as a semi-regular variable based on the short-period variability of $\Delta m<0.5$~mag \citep{Kiss2006}, improving on the previous classifications: long-period variable \citep[LPV;][]{GaiaDR32023, Watson2006, MACHO2008} and nonperiodic variable \citep{ASAS-SN2018}. However, three exceptional events stand out with $\Delta m \sim 1$~mag and a rise time of $\sim$400~d  (Fig.~\ref{fig:fig_dimmings}). We present a comparison of the $V$-band from AAVSO of the Great Dimming of Betelgeuse, the recent dimming of RW Cep \citep[][]{Jones2023, Anugu2023} and the largest dimming $(\Delta V \sim 0.8$~mag) of $\mu$ Cep in the last 50~yr, after the minimum in October 2015. We used the minimum of each event as the zero point for the relative magnitude and date. The first event in [W60]~B90 occurred during the last years of the MACHO survey, between 1999 and 2000, reaching a minimum  $V_{\rm{KC}}=15.4$~mag and abruptly rising $\Delta V_{\rm{KC}}=0.9$~mag after. The next major event is identified in the OGLE $V$-band data between 2011 and 2012, when it suffered another dimming increasing $\Delta V=0.9$~mag, which was remarkably similar in time and brightness to the event in the MACHO data. The last major event occurred between October 2022 and November 2023, with the brightness increasing by $\Delta c\sim1.3$~mag, $\Delta o\sim1$~mag, and $\Delta V\sim1.3$~mag. However, this event has a small discontinuity in the rise, delaying the recovery compared to the other two events. The three events are 4350 and 4250~d apart, corresponding to an average recurrence period of approximately 11.8~years. 

Apart from the three dimming events, we also note a variation in the photometric colors of [W60]~B90 during three different minima in the light curve (Fig. \ref{fig:fig_lightcurve}). In general, a change to a redder color indicates a cooler atmospheric temperature, an increase in extinction due to dust formation, or both. The largest variation was observed by \textit{Gaia} during the minimum in early 2015. $BP-RP$ increased to 3.75~mag, as it faded, but stabilized around 3.45~mag after the recovery. Furthermore, the OGLE data revealed another change in color around 54100~MJD, when a large offset was observed between filters $V$ and $I$. However, the poor sampling of the light curve prevents us from further analyzing this event. Lastly, the ATLAS $c-o$ color increased during the 2022 minimum. Follow-up spectroscopic observations after the minimum revealed a decrease in the \teff~and an enhancement of the extinction in agreement with the changing color (see Sect.~\ref{sec:optical_spectroscopy}). Remarkably, the color changes do not appear at every minimum, which suggests that specific conditions existed during these events.

\subsection{Mid-infrared light curve} \label{sec:mid-IR lightcurve}

\begin{figure}
        \centering
   	\resizebox{1.0\hsize}{!}{\includegraphics{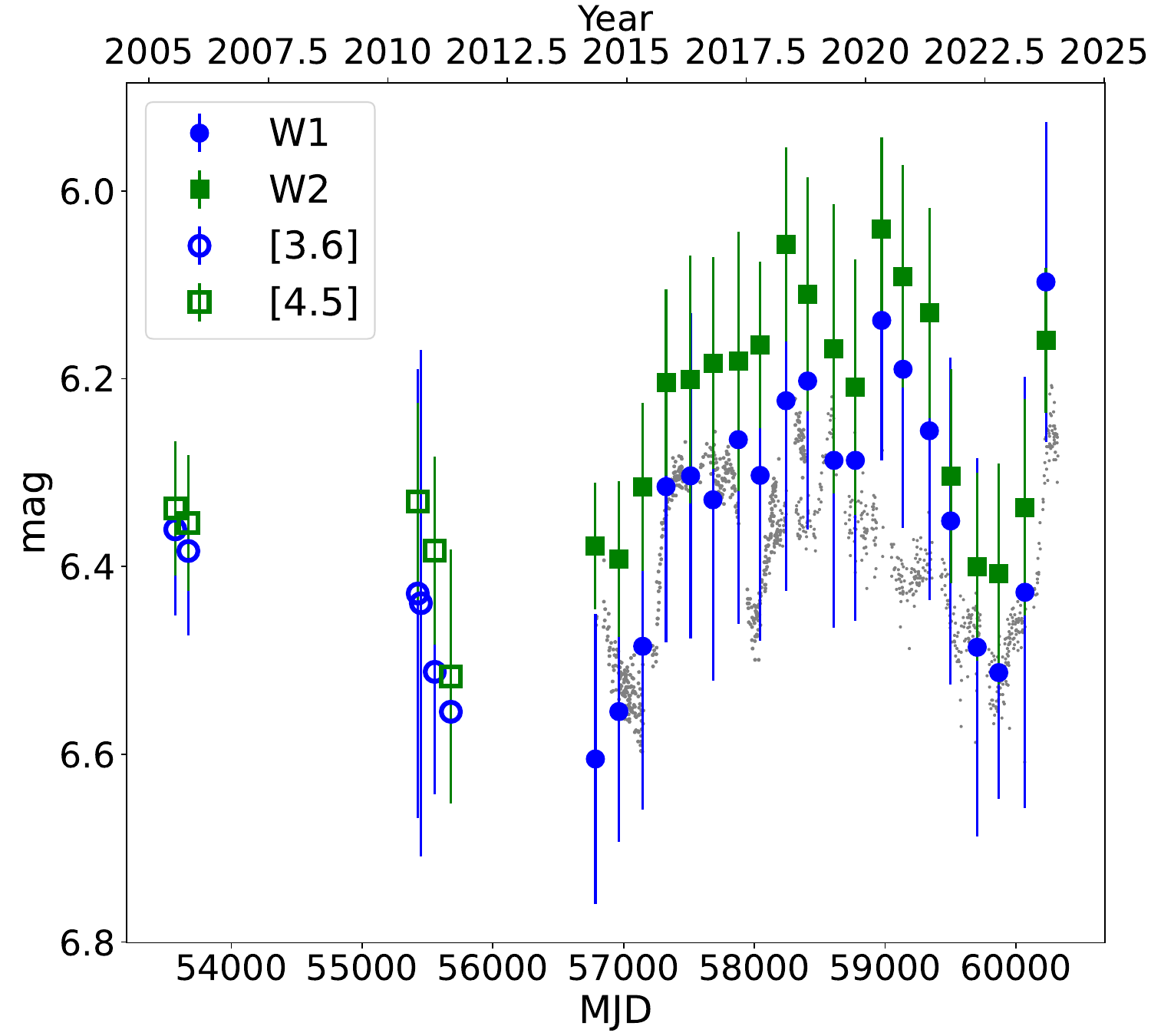}}
  	\caption{Mid-infrared light curve of [W60]~B90 from NEOWISE and {\it Spitzer}. W1 and [3.6] are shown with full and open blue circles, respectively; W2 and [4.5] are shown with full and open green squares. ASAS-SN data are shown in gray with an offset for comparison.}\label{fig:lightcurve_mIR}
\end{figure}

Extreme RSGs do not only show variability in the optical range but may also vary significantly at longer wavelengths. \citet{Yang2018} have found a correlation between the MIR variability, luminosity, and mass-loss rate, with the latter confirmed by \citet{Antoniadis2024}. Therefore, we analyzed the NEOWISE data and compared it with the variability of [W60]~B90 in the optical (Fig. \ref{fig:lightcurve_mIR}). The long-term variability of the data agrees with the optical trend from the ASAS-SN photometry. Contrary to the optical data, no short-term variations were observed. For example, the ASAS-SN data shows a decrease in the brightness at 58000 MJD, which is not detected in the NEOWISE bands. Higher cadence MIR photometry, as well as a comparison of optical to MIR photometry for other RSGs, is needed to confirm whether the short-term variations between the MIR and the optical are indeed decoupled.

The amplitude of the NEOWISE light curve is $\Delta \rm W1=0.51$~mag and $\Delta \rm W2=0.37$~mag. The data cover the last dimming event, yielding an amplitude $\Delta \rm W1=0.42$~mag and $\Delta \rm W2=0.25$~mag. We also calculated the median absolute deviation (MAD), a robust indicator to assess the variability, and found $\rm MAD_{W1}= 0.0903$ and $\rm MAD_{W2}= 0.0918$. Such a large $\rm MAD_{W2}$ places [W60]~B90 in the top four RSGs with the highest $\rm MAD_{W2}$ in the LMC \citep[see Fig. 17 in][]{Yang2018}. Hence, the large MIR variability agrees with the extreme nature of [W60]~B90: one of the most luminous RSG in the LMC with the third highest mass-loss rate for a single RSG \citep{Antoniadis2024}.

\subsection{Periodicity} \label{sec:periodicity}

Periods $P_1=1006$~d and $P_2=453.44\pm0.04$~d have been derived using the MACHO data of [W60]~B90 \citep{Groenewegen2009, Groenewegen2018}. Note that other studies reported $P=776$~d (with $\Delta I_c=0.39$~mag) from AAVSO photometry, and a long secondary period (LSP) of $4900$~d, $\Delta m=0.12$~mag from 50~yr of the digitized Harvard Astronomical Plate Collection  \citep{Watson2006, Chatys2019}.

We used the period$-$luminosity (P-L) relations in the NIR and MIR \citep{Yang2011} to calculate the predicted period and compare it with $P_1$. We compiled the photometry of [W60]~B90 in the bands $J$, $H$, $K_s$, [3.6], and [4.5] from the catalog of RSGs in the LMC published by \citet{Yang2018}. We calculated the periods to be 829, 929, 1032, 1107, and 1041~d from $J\!=\!8.37$, $H\!=\!7.40$, $K_s\!=\!6.83$, [3.6] $=\!6.29$, and [4.5] $=\!6.29$~mag, respectively. Despite the lack of error bars in $P_1=1006$~d, we consider the derived periods from $H$, $K_s$, [3.6], and [4.5] consistent with the expected. In fact, \citet{Yang2011} mentioned that $K_s$ and [3.6] are the most reliable bands to use their P-L relations as they are the least affected by extinction and provide the tightest relations. As they did not de-redden the photometry, RSGs with high $A_V$ might differ from their predicted periods in the bands affected by extinction. Therefore, we attribute the shorter period from the $J$-band to the high extinction $A_V>3$~mag reported for [W60]~B90 (see Sect.~\ref{sec:optical_spectroscopy}).   

%------------------------------------------------------------------
%------------------------------------------------------------------
%------------------------------------------------------------------

\subsection{Optical spectroscopy} \label{sec:optical_spectroscopy}

\begin{figure*}
        \resizebox{1.0\hsize}{!}{\includegraphics{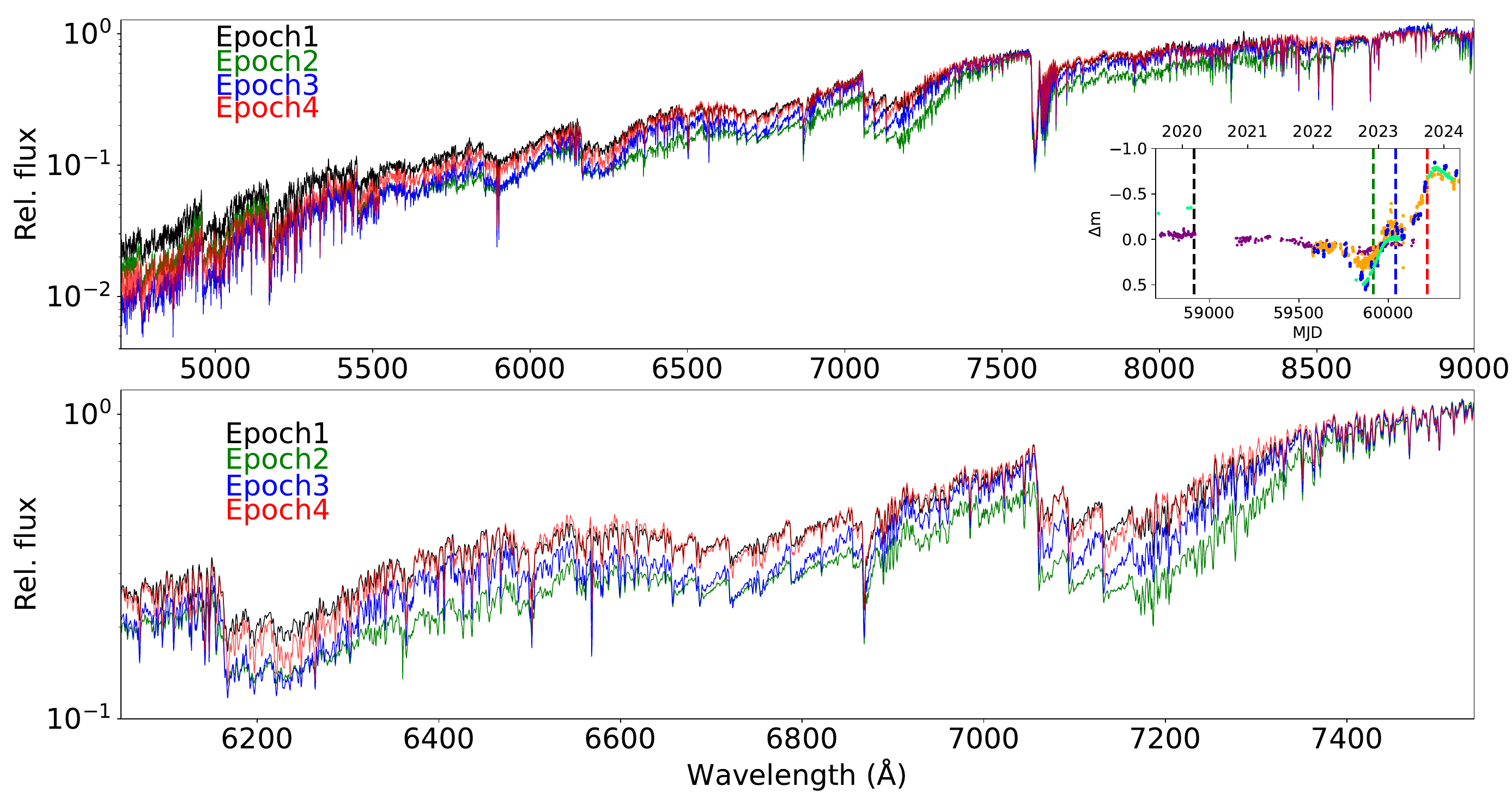}}
  	\caption{Comparison of the spectroscopic epochs of W60~[B90] in the optical (\textit{top}) and the main TiO bands at 6150 and 7050~\r{A} (\textit{bottom}). The inset shows the light curve (as in Fig. \ref{fig:fig_lightcurve}) indicating each spectroscopic epoch with the dashed lines, following the color code of the spectra.}\label{fig:fig_spec_comparative}
\end{figure*}

\begin{figure*}
        \resizebox{1.0\hsize}{!}{\includegraphics{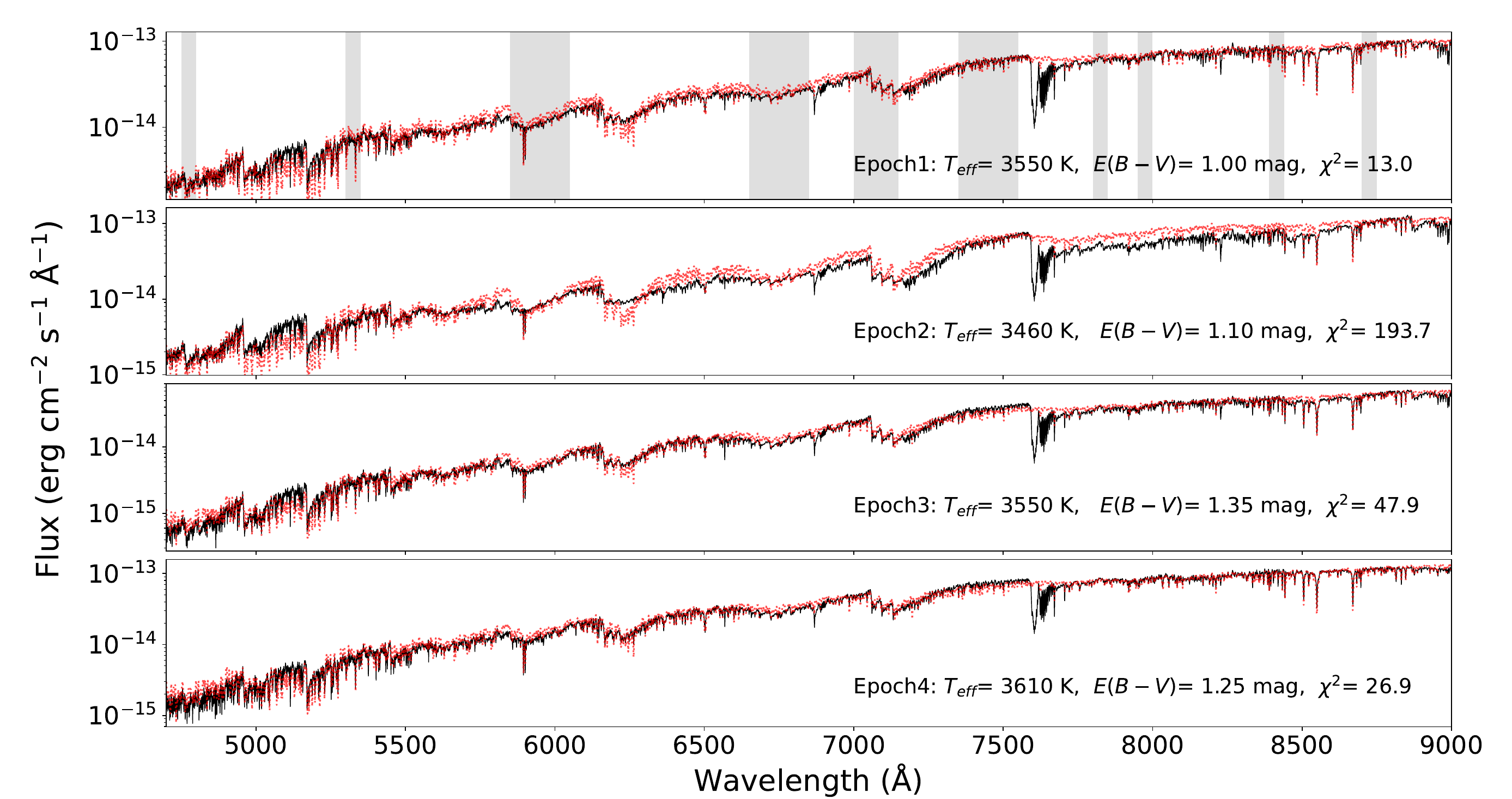}}
  	\caption{Best \textsc{marcs} model fit (red) for each epoch from MagE (black). Shaded areas in Epoch 1 show the spectral regions used in the fitting.}\label{fig:fig_MARCS}
\end{figure*}

We compared the optical spectra of the four epochs presented in Fig. \ref{fig:fig_spec_comparative} to analyze the evolution of the star with time. We detect changes in the shape of the SED and the depth of the TiO bands throughout the epochs, implying a spectral-type variation from M3~I to M4~I. Changes in the SED imply either a variation in the \teff, $E(B-V)$, or both. Since the metallicity $Z$ of the star does not change, the strength of the TiO bands is exclusively related to the \teff\, \citep[e.g.,][]{Levesque2007} and the wind of the star \citep{Davies2021}, being deeper when the \teff~is cooler and the wind is stronger. We used the grid of \textsc{marcs} models described in \citet{deWit2023} to obtain the physical parameters for each spectral epoch. We primarily fit for \teff~and $E(B-V)$, keeping $Z$ and $\log(g)$ fixed from the results in the $J$-band (see Sect. \ref{sec:nIR_spectroscopy}). We derived $Z=-0.25$~dex and $\log(g)=+0.5$~dex from the local thermodynamic equilibrium (LTE) \textsc{marcs} models, in contrast to \citet{deWit2023}, who assumed a LMC-like metallicity ($Z=-0.38$~dex) and obtained $\log(g)=-0.2$~dex from the Ca~\textsc{ii} triplet. The discrepancy in $\log(g)$ does not affect our fit in the optical as the TiO bands are temperature indicators and not sensitive to gravity. Finally, we performed the fitting in two stages. In the first iteration, we fitted spectral regions including shorter wavelengths to constrain the shape of the SED and get more accurate values for $E(B-V)$. Then, fixing $E(B-V)$ in the second iteration, we only use spectral regions affected by the TiO bands to obtain the \teff~of the star. The best-fit models to each epoch are shown in Fig. \ref{fig:fig_MARCS} and their parameters are shown in Table~\ref{tab:MARCS_params}.
 
We obtained Epoch1 at the beginning of 2020 when the RSG exhibited low-amplitude variability (see Fig~\ref{fig:fig_lightcurve}). Indeed, the fit of Epoch1 reveals the lowest $\chi^2$ among our sample, implying minor deviations from the model spectra, suggesting the RSG atmosphere was stable. During the following years (2020-2022), the brightness decreased $\Delta g = 0.3$~mag in ASAS-SN, although the real change in brightness might be larger (see discussion in Sect.~\ref{sec:opt_lightcurve}). We obtained Epoch2 two months after the photometric minimum. This spectrum reveals the strongest TiO bands of all epochs, implying either a decrease in the \teff, an increase of \Mdot, or a combination of both. The spectrum suggests a complex atmospheric structure following the photometric minimum, which cannot be reproduced by a single \textsc{marcs} model (see Sect. \ref{sec:epoch2}), and we consider the \teff~and the $E(B-V)$ derived for Epoch2 to be unreliable. We obtained Epoch3 four months later, when the RSG exhibited a plateau during the recovery, after initially increasing $\Delta o=0.5$ mag compared to the minimum. It shows weaker TiO bands than Epoch2, and we found a higher best-fit \teff~according to this change. However, $E(B-V)$ is considerably higher than the previous epochs, implying that the extinction in the line of sight increased after the minimum. Finally, we obtained Epoch4 two months before the maximum in late 2023. In this epoch, it was brighter than Epoch2 by $\Delta o=0.9$~mag, and the RSG exhibited the highest measured \teff. The SED is not as extinct as in Epoch3 but is steeper than in the first two spectra, in agreement with the derived $E(B-V)$. 

\begin{table*}[!htb]
\caption{Physical parameters of [W60] B90 from spectroscopy} \label{tab:MARCS_params}
\smallskip 
\small
\begin{center}
%\begin{tabular*}{l c c c c r}
\renewcommand{\arraystretch}{1.3}
\begin{tabular*}{0.6\textwidth}{@{\extracolsep{\fill}}lccllr}
\hline \hline \\[-10pt]
\multicolumn{6}{c}{Atomic lines in the $J$-band$^a$} \\
\\[-10pt] \hline
  Name   & Model & $Z$ & $T_{\rm eff, J}$ & log($g$)  & $\chi^2$  \\
  & & (dex) & (K) & (dex) &   \\
\hline
  \multirow{2}{*}{EpochJ} & \textsc{marcs} LTE & $-0.25\substack{+0.25 \\ -0.12}$ & $3970\substack{+130 \\ -280}$ & $-0.20\substack{+0.20 \\ -0.30}$ &  63.1   \\ 
   & \textsc{marcs} NLTE & $+0.00\substack{+0.20 \\ -0.10}$ & $3900\substack{+150 \\ -100}$ & $+0.50\substack{+0.00 \\ -0.75}$ &  36.4   \\
%\bottomrule
\\[-5pt]
\hline \hline \\[-10pt]
\end{tabular*}
\begin{tabular*}{0.6\textwidth}{@{\extracolsep{\fill}}lcclllr}
\multicolumn{7}{c}{TiO bands from the optical} \\
\\[-10pt] \hline
    & Spectral type & ATLAS $o$ & $T_{\rm eff, TiO}$ & $E(B-V)$  & ${A_V}^b$ & $\chi^2$ \\ %& ASAS-SN-$V$ \\
     & & (mag) & (K) & (mag) & (mag) &  \\
\hline
  Epoch1$^c$ & M3~I & --  & 3550$\pm$40  & 1.00$\pm$0.15 & 3.41$\pm$0.51 & 13.0  \\ %& 13.91$\pm$0.03
  Epoch2 & M4~I & 12.6$\pm$0.1 & 3460$^{+20}_{-30}$  & 1.10$\pm$0.10 & 3.75$\pm$0.34 & 193.7   \\ %& 14.10$\pm$0.01
  Epoch3 & M3~I & 12.3$\pm$0.1 & 3550$^{+40}_{-30}$  & 1.35$^{+0.10}_{-0.05}$ & 4.60$^{+0.34}_{-0.17}$ & 47.9   \\ %& 14.06$\pm$0.02
  Epoch4 & M3~I &  11.8$\pm$0.1 & 3610$^{+60}_{-50}$  & 1.25$^{+0.10}_{-0.05}$ & 4.26$^{+0.34}_{-0.17}$ & 26.9   \\ %13.86$\pm$0.02 &
\hline
\end{tabular*}
\tablefoot{$^{(a)}$ Assuming $Z=-0.25$~dex and $\log g=-0.2$~dex from the $J$-band fit.$^{(b)}$ Converted from $E(B-V)$ assuming $R_V=3.41$. $^{(c)}$ \teff=3570$^{+60}_{-50}$~K and $E(B-V)=1.00\pm0.14$~mag from \cite{deWit2023}.}
    
\end{center}
\end{table*}

\subsection{Spectroscopy during the dimming (Epoch2)}\label{sec:epoch2}
\begin{figure*}
        \centering
        \resizebox{1.0\hsize}{!}{\includegraphics{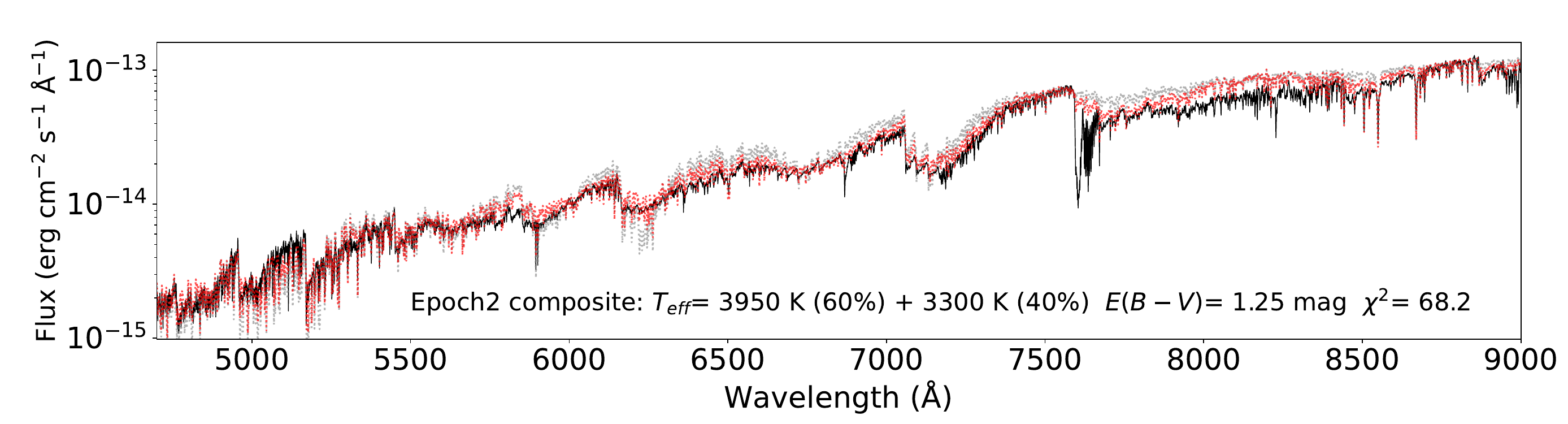}}
  	\caption{Comparison between the best composite model (red), with 60\% of the flux from a single $T_{\rm eff1}=3950$~K and 40\% from $T_{\rm eff2}=3300$~K, and the best single model (gray) from Fig. \ref{fig:fig_MARCS} for Epoch 2.}\label{fig:fig_MARCS_composite}
\end{figure*}

The Great Dimming of Betelgeuse was explained by a clump of dust in the line of sight or, alternatively, a cold patch in the atmosphere \citep{Montarges2021}. We attempted to model the latter by creating a grid of composite \textsc{marcs} models. We created composite models from the superposition of two single models ($T_{\rm eff1}$ and  $T_{\rm eff2}$) with weighted fluxes from each model in steps of 20\% (e.g., 80-20\% or 60-40\%). We used \teff~from 3300--4500~K in steps of 50~K and assumed $Z=-0.25$~dex and $\log(g)=-0.2$~dex as in Sect. \ref{sec:optical_spectroscopy}. 

We obtained the best fit for $60\%$ of $T_{\rm eff1}=3300$~K, $40\%$ of $T_{\rm eff2}=4500$~K, $E(B-V)=1.2$~mag, and $\chi^2=62.4$. The composite model considerably improves the fit, but the temperatures found are at the edges of the grid. Fixing T$_{\rm eff1}=3950~K$ using the result of EpochJ (see Sect.~\ref{sec:nIR_spectroscopy}), we derived the best fit for $60\%$ of $T_{\rm eff1}=3950$~K to be $40\%$ of $T_{\rm eff2}=3300$~K and $E(B-V)=1.25$~mag, with a $\chi^2=68.2$ (Fig. \ref{fig:fig_MARCS_composite}), again finding a $T_{\rm eff2}$ at the edge of the grid. Nevertheless, we argue that the \textsc{marcs} models cannot always reproduce the real photosphere of an extreme RSG. Physical processes such as 3D assumptions, wind, or magnetic fields, are currently missing in the \textsc{marcs} models recipes. Hence, a new generation of models including these processes is needed to properly reproduce the complete nature of RSGs. 

\subsection{Near-infrared spectroscopy} \label{sec:nIR_spectroscopy}
\begin{figure*}
        \centering
        \resizebox{1.0\hsize}{!}{\includegraphics{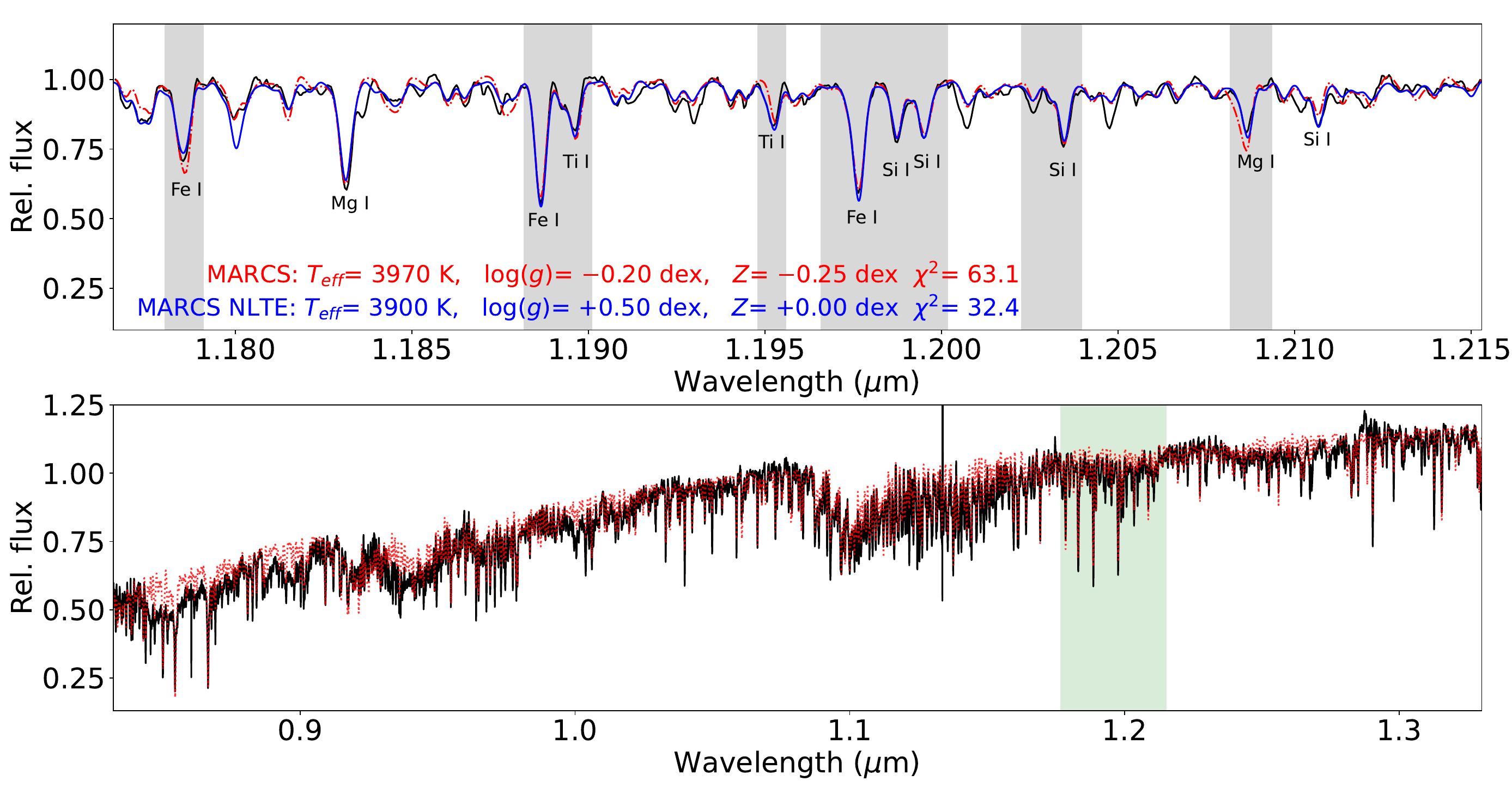}}
  	\caption{Derived parameters from the FIRE spectrum. \textit{Top}: Best fit of the FIRE spectrum (solid black) in the $J$-band for the LTE \textsc{marcs} models (dashed-dot red) and the NLTE-corrected version (solid blue) with the NLTE effects applied to the indicated lines. The spectral regions used in the fitting are shown with gray shades. \textit{Bottom}: Best LTE \textsc{marcs} model (dotted red) from the $J$-band to the FIRE spectrum (black). The model was reddened with $E(B-V)=1.05$~mag to match the SED. The green-shaded region highlights the spectral region shown in the upper plot.}\label{fig:fig_MARCSnIR}
\end{figure*}

We applied the nonLTE (NLTE) \textsc{marcs} models \citep{Bergemann2012, Bergemann2013, Bergemann2015} to fit the atomic lines in the $J$-band of our FIRE spectrum. We created a grid of models with a range \teff= 3300$-$4500~K in steps of 25~K, $\log(g)=-0.5$, $+0.5$~dex in steps of 0.25~dex, microturbulent velocities from 2.5 to 5.5~km~s$^{-1}$ in steps of 0.5~km~s$^{-1}$ and $Z=-0.38, -0.25, -0.1, +0.0$ and $+0.2$~dex. Also, we used the LTE grid presented in the previous section to compare the results from the NLTE models with the canonical \textsc{marcs} models. We found the best-fit model with reduced $\chi^2$ and the uncertainties from the 1$\sigma$ interval of the distribution. We followed the approach presented in \citet{Patrick2017} and fitted the lines Fe~\textsc{i} $\lambda\lambda$1.178327, 1.188285, 1.197305, Si~\textsc{i} $\lambda\lambda$1.198419, 1.199157, 1.203151, Mg~\textsc{i} $\lambda\lambda$1.208335 and
Ti~\textsc{i} $\lambda\lambda$1.189289, 1.194954. We excluded the line Mg~\textsc{i} $\lambda\lambda$1.182819 because the feature on the right wing of the line profile is not in the models and could compromise the diagnostic of the line. Similarly, we rejected Si~\textsc{i} $\lambda\lambda$1.210353 due to the difficulty of establishing the continuum level during the fit. Nevertheless, both lines are satisfactorily reproduced by the best-fit model probing that their rejection did not compromise the result. 

The results presented in Fig. \ref{fig:fig_MARCSnIR} and Table \ref{tab:MARCS_params} show a strong disagreement of $\gtrsim$300~K between the \teff~derived from the TiO bands ($T_{\rm{eff,TiO}}$) in the optical spectra and the atomic lines fit in the $J$-band ($T_{\rm{eff,J}}$). The NLTE models constrain the $T_{\rm{eff,J}}$ better than the LTE models, although both results are consistent within the errors. Both indicate higher $Z$ than the commonly $Z=-0.38$~dex assumed for the LMC, with the LTE models suggesting a slightly higher $Z$ than the LMC, while the NLTE models favor a solar $Z$. The $\log(g)$ results are poorly constrained as the errors span over the whole range of the grid. The best NLTE model finds $\log(g)=+0.50\substack{+0.00 \\ -0.75}$~dex, similar to what \citet{deWit2023} derived from the Ca~\textsc{ii} triplet in Epoch1 and consistent within error with the LTE result $\log(g)=-0.2\substack{+0.20 \\ -0.30}$~dex as the best solution. We applied $\log(g)$ and $Z$ from the LTE result to constrain the physical parameters in the optical for consistency, as the TiO bands do not have NLTE correction (see Sect.~\ref{sec:optical_spectroscopy}). Therefore, using $Z=+0.00$ from the NLTE in the optical instead of $Z=-0.25$ would lead to an overestimation of the abundances, affecting the $T_{\rm{eff,TiO}}$. However, the discrepancy in $\log(g)$ between the LTE and the NLTE result is negligible as the TiO bands are not sensitive to this parameter.

%------------------------------------------------------------------
%------------------------------------------------------------------
%------------------------------------------------------------------
\section{Discussion}\label{sec:discusion_gen}

\subsection{Bow shock, bar and runaway status}\label{sec:discusion_bow_shock}

Only three single Galactic RSGs with a bow shock have been identified so far: Betelgeuse, IRC-10414, and $\mu$~Cep, with only the bow shock of IRC-10414 being visually detected in the optical range. An analysis of [S~\textsc{ii}]/H$\alpha$ revealed a ratio of 0.3 \citep{Gvaramadze2014}, which is even lower than the values reported around [W60]~B90. Among the three RSGs, only Betelgeuse exhibits a bar, which is located at 0.5~pc from the star \citep[assuming a distance of 200~pc;][]{Harper2008, Harper2017} and which remains unexplained. It might be the relic of the blue supergiant (BSG) wind interaction with the ISM, just before Betelgeuse recently became a RSG \citep{Mackey2012}. Although new constraints on the evolutionary status of Betelgeuse reject this hypothesis, and an interstellar origin was proposed instead \citep{Decin2012, Meyer2021}. The bar could be the edge of an interstellar cloud illuminated by Betelgeuse or a linear filament in the interstellar cirrus. Some efforts have been made to detect the bar and the bow shock in the optical with no success. The glow prevented the detection of the bar through imaging even setting the star outside the field of view (private communication with Dr. Jonathan Mackey). In the case of [W60]~B90, the bar structure at 1~pc is partially shocked, but only in the southeastern part. The movement of [W60]~B90 towards the bar and the shocked material found between them and in the bar confirms a causal connection. However, the interpretation of our findings requires further work and is beyond the scope of this paper.

[W60]~B90 moves towards the bar with a peculiar velocity between $16-25 (\pm11)$~km~s$^{-1}$ (Table~\ref{sec:apendix_PM_values}.1). Given that hydrogen is the most abundant element in the gas, we can assume that it is a good tracer of the CSM and use the $\sim$10~km~s$^{-1}$ difference in RV between H$\alpha$ and the star to construct the 3D velocity. Therefore, [W60]~B90 moves with a $19-27(\pm11)$~km~s$^{-1}$ velocity, establishing it as a walkaway star, on the brink of the runaway limit \citep[>30~km~s$^{-1}$;][]{Renzo2019}. The speed of sound in the low-density isothermal warm neutral medium is on the order of $\sim1$~km~s$^{-1}$ \citep{Cox2012}. Therefore, the RSG would move supersonically in the medium even considering the 3D lower limit of 8~km~s$^{-1}$. Studies of the case of Betelgeuse demonstrated that lower velocities than 50~km~s$^{-1}$ can produce an observable bow shock during $\sim$100~kyr, a significant fraction of the post-main sequence evolution  \citep{Mackey2012}. Also, bow shocks with low stellar velocities form clumpy substructures due to Kelvin-Helmholtz instabilities, which could explain why we do not measure homogeneous [S~\textsc{ii}]/H$\alpha$ values around the star \citep{Mohamed2012}.  

Further releases from \textit{Gaia} will improve the uncertainties on the peculiar velocity, allowing us to trace back its movement, discern its birthplace, and speculate the cause of its ejection, either by dynamical interactions or from a SN kick \citep{Stoop2023}. 

Apart from detecting shocked material where the bow shock is expected, we also find enhanced [S~\textsc{ii}]/H$\alpha$ at northern positions close to the star. This might indicate an inhomogeneous CSM and clumpy, asymmetric mass-loss events. Observations in the MIR have shown multiple arcs and similarly strong asymmetries in the CSM of Betelgeuse. \citet{Decin2012} demonstrated that a combination of anisotropic mass-loss processes and the influence of galactic magnetic fields might explain the multiple arcs and clumps around Betelgeuse. Further observations are needed to spatially resolve the circumstellar environment of [W60]~B90. Identifying the hypothesized bow shock or inhomogeneous structures around the star would help to constrain the recent mass-losing history of [W60]~B90.

 \subsection{Properties of the dimming events}\label{sec:discusion_phot_var}

Our spectroscopic analysis revealed physical properties similar to those of Betelgeuse during the Great Dimming (see Sect. \ref{sec:epoch2} and \ref{sec:discusion_spec_var}), therefore, we speculate that the same physical mechanism drives both events. \citet{MacLeod2023} explained the Great Dimming as a result of a hot convective plume that forms in the turbulent envelope, breaking free from the surface and triggering a mass ejection. However, the Great Dimming displayed a more abrupt brightness rise on a shorter timescale ($\sim$200~d) than the rise of the events of our RSG ($\sim$400~d). The radius of [W60]~B90 is $\sim$1200~\rsun, while that of Betelgeuse is reported to be between 750--1000~\rsun~\citep{Joyce2020, Kravchenko2021}. Contrarily, $\mu$~Cep \citep[1259~\rsun;][]{Josselin2007} and the Galactic hypergiant RW Cep \citep[900–1760~\rsun;][]{Anugu2023} have a comparable radius to [W60]~B90, and both stars exhibit a similar rise after the minima (Fig.~\ref{fig:fig_dimmings}). The large uncertainty in the radius of RW~Cep derives from the large uncertainty in the distance and, hence, luminosity. We speculate that the timescale of these events is related to the radius of the stars, as a more extended atmosphere needs more time to stabilize. A similar idea was already presented in the analysis of the dimming of RW~Cep \citep{Anugu2023}, but it needs to be tested with more dimming events in RSGs. However, assuming a similar timescale in the recovery of RW~Cep with respect to $\mu$~Cep and [W60]~B90, would constrain the radius of RW~Cep to $\sim$1200~\rsun. The dimming events of [W60]~B90 have a recurrence of $\sim$11.8~yr, while the dimming of $\mu$ Cep is unique over the past 50~yr and the Great Dimming of Betelgeuse is unique over the last 100~yr. Only one dimming event has been observed for RW Cep, but the dust shells detected around the hypergiant suggest that it may have undergone several mass ejections over the last century \citep{Anugu2023, Jones2023}. Moreover, the rising plume on Betelgeuse disturbed its pulsation period, switching from the $\sim$400~d fundamental period to the $\sim$200~d overtone \citep{MacLeod2023}. We do not find a change in the periodicity in any of the three dimmings. 

VY~CMa is another red hypergiant that experienced several dimming events during the last century. The most extreme event occurred in the 90s when the star decreased the brightness by $\sim 3$~mag. The other events showed variations of $\Delta m \sim 1.5$~mag \citep{Humphreys2020} which are still larger than the events of [W60]~B90 ($\Delta m \sim 1$~mag). Moreover, the $\sim$500~d rise time on the VY~CMa is larger than the $\sim$400~d of [W60]~B90, which agrees with VY~CMa having a bigger size \citep[$1420\pm120$~\rsun;][]{Wittkowski2012} than [W60]~B90 ($\sim$1200~\rsun). Different gaseous knots, arcs, and irregular structures surrounding VY~CMa have been identified as discrete mass ejections related to each minimum during the last century. These massive gaseous outflows explain the high mass loss of VY~CMa \citep{Humphreys2005, Humphreys2007, Humphreys2020}. On the other hand, although similar structures have been identified around Betelgeuse, they do not explain the overall mass-loss rate of this RSG but contribute to it \citep{Humphreys2024}. Spatially resolving the CSM of [W60]~B90 could reveal the presence of gaseous structures, show if there is a correlation with the 11.8~yr recurrence of the dimmings, and if they explain the mass-loss rate as in VY CMa or only contribute to it as in Betelgeuse. Extending the long-term variability study to other luminous RSGs is needed to understand how common such events are, and confirm their dependence on the size of the RSG.

The fundamental period $P_1=1006$~d \citep{Groenewegen2018} and the 4900~d LSP of [W60]~B90 are considerably larger than the fundamental $P\sim400$~d \citep{Kiss2006} and the 2000-2365~d LSP \citep{Chatys2019,Joyce2020} of Betelgeuse. They are also consistent with the expected $P$ from the P-L relations and the higher luminosity of [W60]~B90. Although \citet{Chatys2019} do not provide errors, their 13.5~yr LSP of [W60]~B90 is suspiciously close to the 11.8~yr of the dimming recurrence found in this paper. Furthermore, the periods found in [W60]~B90 are remarkably similar to the fundamental period P=880~d and LSP of 4400~d of $\mu$~Cep. Although this RSG is reported to be comparable in luminosity and radius to [W60]~B90 (see Sect.~\ref{sec:dis_mdot}), no recurrence of dimming events has been reported in $\mu$~Cep to date.

\subsection{Spectral variability}\label{sec:discusion_spec_var}

The TiO bands are the primary spectroscopic feature for classifying M-stars in the optical. \citet{Dorda2016} found that $30\%$ of the $\sim$500 RSGs in their LMC sample showed spectral type variability with a mean change of two spectral subtypes. They also reported that cool RSGs are more likely to exhibit spectral-type changes. In the case of [W60]~B90, we found the depth of the TiO bands to vary among the epochs, yielding a spectral type of M3~I in Epoch1 \citep{deWit2023} and M4~I in Epoch2. However, observations of the star during its maximum and minimum would likely yield larger spectral variability. These results demonstrate the uncertainty of spectroscopically classifying variable RSGs based on a single observation. 

A tomographic analysis of $\mu$~Cep and Betelgeuse explained the correlation between $T_{\rm{eff,TiO}}$ and the optical variability as an effect of the convective cells in the atmosphere \citep{Kravchenko2019, Kravchenko2021}. The $T_{\rm{eff,TiO}}$ and the brightness decrease as the material rises, while both increase once it falls, creating a hysteresis loop between the RV of the atomic lines and the $T_{\rm{eff,TiO}}$. We report a similar trend between $T_{\rm{eff,TiO}}$ and the optical variability for [W60]~B90 during the brightening in 2022. The Epoch2 spectrum was obtained 2 months after the minimum in 2022, showing spectral features incompatible with single \textsc{marcs} models. \citet{Montarges2021} explained the Great Dimming as a large cold spot rising in the atmosphere of Betelgeuse. Big convective cells at different temperatures can create large cold spots that cause deeper TiO bands, decreasing the brightness in the optical. Furthermore, if convection is strong enough, it can result in a mass ejection \citep{MacLeod2023, Drevon2024}. This scenario is consistent with the enhanced extinction reported in the epochs following the minimum Epoch3 and Epoch4, where $A_V$ changes from $3.41\pm0.51$~mag before the event to $4.60^{+0.34}_{-0.17}$~mag several months after. This demonstrates the importance of using the light curve to interpret whether the RSG was observed during a stable state, or during a brief minimum or maximum. Further investigation is needed to reveal the origin of the hysteresis loops and how they are related to episodic mass-loss events. 

\subsection{\texorpdfstring{$T_{\rm{eff,TiO}}$}{Teff,TiO} versus \texorpdfstring{$T_{\rm{eff,J}}$}{Teff,J}}\label{sec:discusion_teff_var}

We found a strong discrepancy in \teff~when we obtained the physical parameters from different spectral ranges. While the TiO bands in the optical suggest $T_{\rm{eff,TiO}}\approx3550$~K, we derived $T_{\rm{eff,J}}=3900$~K from the atomic lines in the $J$-band. \citet{Davies2013, Davies2015} already reported a discrepancy in \teff~derived from the TiO bands compared with the $J$-band or the SED fits, which is related to the formation zone of each diagnostic. However, if the \textsc{marcs} models were consistent, one single model would describe the atmosphere of the star with one single \teff. Given these discrepancies, \citet{deWit2024} derived a \teff~scaling relation based on the \textsc{marcs} models to scale $T_{\rm{eff,TiO}}$ to a more secure \teff. Using their relation, we find $T_{\rm{eff,J}}=3960$~K, which is consistent with our results. 

Previous studies on luminous RSGs have also reported very cool $T_{\rm{eff,TiO}}$, which cannot be reproduced by the theoretical evolutionary models \citep[e.g.,][]{Levesque2007, deWit2024}. \citet{Davies2021} demonstrated that adding a $\dot{M}$ to the \textsc{marcs} models results in an enhancement in the TiO band strengths, similar to the effect of decreasing the \teff. The $T_{\rm{eff,TiO}}$, therefore, might be underestimated in evolved luminous RSGs with strong $\dot{M}$ as the \textsc{marcs} models do not account for it. Recently, by implementing $\dot{M}$ to the \textsc{marcs} models, \citet{GonzalezTora2024} were able to reconcile spectral features in the near and MIR that did not match with the canonical \textsc{marcs} models. Our results reinforce the urgent need for a complete grid of models with $\dot{M}$ to break the degeneracy in the optical and reconcile the $T_{\rm{eff,TiO}}$, the $T_{\rm{eff,J}}$ and the evolutionary models. Extending the 1D LTE assumptions from the \textsc{marcs} models to 3D magneto-hydrodynamic ones will also improve the modeling of the strong convection in RSGs \citep[e.g.,][]{Kravchenko2019,Ma2024}

Furthermore, the addition of NLTE corrections to the \textsc{marcs} models for the atomic lines in the $J$-band considerably impacts the results. It decreases the uncertainty in \teff~and it reveals a high metallicity ($Z=+0.00\substack{+0.20 \\ -0.10}$~dex), which is inconsistent with the mean metallicity of the LMC \citep[$Z=-0.37\pm0.14$~dex;][]{Davies2015}. Although it is more metal-rich than expected for the LMC, [W60]~B90 is not the first cool supergiant with solar-like $Z$, as six more cases are already known in the LMC \citep{Tabernero2018}. The solar $Z$, however, is in conflict with the low-$Z$ environment suggested by the faint nitrogen nebular emission. The enhanced $Z$ can be explained by extra rotational mixing occurring in its interior during its evolution enhancing the metal content in the surface, by a binary history with mass-transfer changing the abundances, or even a merger with a companion.  

\subsection{Binarity} 
\label{sec:binary}

Recent works have estimated the fraction of RSGs in binary systems to be at least 15\% \citep{Dorda2021, Patrick2022}. We explored the RV variations from the Ca~\textsc{ii} triplet (Table~\ref{tab:rad_vel}) and found all epochs to be consistent with the \textit{Gaia} DR3 value, except for Epoch3. We cannot attribute the difference of less than 10~km~s$^{-1}$ to the presence of a companion, as tomographic studies have revealed that such variations can be explained by the rising and falling of material in the atmosphere \citep{Kravchenko2019, Kravchenko2021}. A pilot study by \citet{Patrick2020} suggested a correlation between the RV variations and the luminosity of the RSGs, which might be connected to stronger hysteresis loops. Since [W60]~B90 is close to the observed upper luminosity of RSGs in the LMC \citep{Davies2018}, it is expected to have larger variability and RV differences. Therefore, we argue that the small discrepancy in the RV is a combined effect of the changes in the atmosphere due to its evolutionary status and the intrinsic error of $\pm7$~km~s$^{-1}$ from the observations. Moreover, \textit{Gaia}~DR3 indicators such as the RUWE, astrometric excess noise, and binary probability parameter do not support a binary scenario. We also report the absence of a counterpart in \textit{Swift} observations and the nondetection of blue excess in the SED, disfavouring a hot companion. Therefore, we conclude that [W60]~B90 is currently a single star given the total lack of evidence for a companion.

\begin{table}[!htb]
\caption{Radial velocity of Ca~\rm{II} triplet} \label{tab:rad_vel} 
\small
\renewcommand{\arraystretch}{1.4}

\centering
\begin{tabular*}{0.45\columnwidth}{c|c}
  \hline \hline
  Spectrum & RV \\
   & (km s$^{-1}$) \\
  \hline
  Epoch1 & $264\pm4$ \\
  EpochJ & $268\pm2$ \\
  Epoch2 & $261\pm3$ \\
  Epoch3 & $272\pm2$ \\ 
  Epoch4 & $266\pm3$ \\
  \hline
  \textit{Gaia} DR3 & $263.49\pm1.02$ \\
  \hline
\end{tabular*}
\tablefoot{Spectral resolution error is $\pm7$ km s$^{-1}$.}\\
\end{table}

\subsection{Mass-loss rate}\label{sec:dis_mdot}

\begin{figure}
        \centering
   	\resizebox{1.0\hsize}{!}{\includegraphics{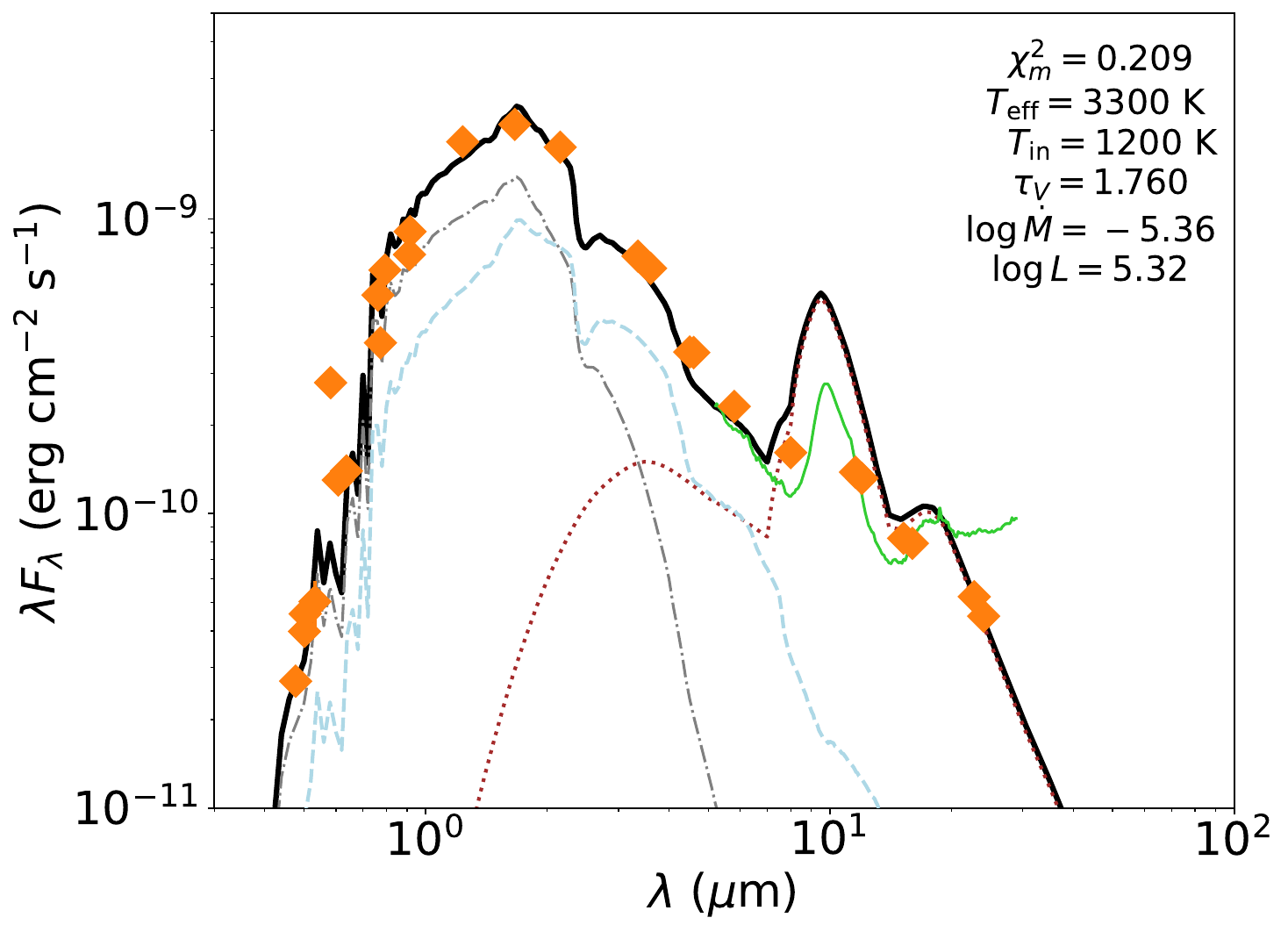}}
  	\caption{SED of [W60]~B90. The orange
diamonds show the observations and the black line is the best-fit model from \texttt{DUSTY}, which is a superposition of attenuated flux (dashed light blue),
the scattered flux (dot-dashed gray), and dust emission (dotted brown). The green curve represents the \textit{Spitzer} IRS spectrum.}\label{fig:fig_Mdot}
\end{figure}

\citet{Antoniadis2024} determined the mass-loss rates of over 2000 RSGs in the LMC using the radiative transfer code \texttt{DUSTY}, finding $\dot{M}=5.1\substack{+5.1 \\ -1.7}\times10^{-6}~$\msun~yr$^{-1}$ for [W60]~B90. In this work, we recomputed the \Mdot~with identical assumptions, but including synthetic photometry from the IRS spectrum (see Sect.~\ref{sec:data_spit}) to improve the SED fitting. We present the new fit in Fig.~\ref{fig:fig_Mdot}, which results in $\dot{M}=4.4\substack{+5.1 \\ -1.7}\times10^{-6}~$\msun~yr$^{-1}$ with a best-fit optical depth $\tau_V=1.76$ for the best fit of \teff~$=3300$~K and inner dust shell temperature $T_\mathrm{in}=1200$~K \citep[for the description of the fitted parameters see][]{Antoniadis2024}. This \Mdot~makes [W60]~B90 the third highest mass-losing probably single RSG in the \citet{Antoniadis2024} sample. Moreover, it is the second highest $\dot{M}_{\rm dust}$ among the oxygen-rich stars in the study of evolved stars in the LMC from \citet{Riebel2012}. These results underline the extreme nature of [W60]~B90 and agree with it being one of the most variable RSGs in the MIR in the LMC \citep{Yang2018}. 

From the properties of a bow shock, one can derive the mass-loss rate of the producer. Therefore, we compare the mass-loss rate and the general properties of [W60]~B90 with the three known, Galactic RSGs with a bow shock in Table \ref{tab:comparative}. All of them have considerable mass-loss (\Mdot~$>10^{-6}$ \msun~yr$^{-1}$) and high luminosity ($\log(L/\lsun)>5.0$~dex). IRC-10414 and $\mu$~Cep are fast runaways, while Betelgeuse is at the limit between a walkaway and a runaway star. Remarkably, Betelgeuse is the most compact of the RSGs, exhibiting shorter periods, while [W60]~B90 and $\mu$~Cep have comparable periodicity (see Sect.~\ref{sec:discusion_phot_var}). However, none of the physical properties stand out as a common signature of a bow shock. External factors to the RSGs such as their environment or the specific evolution likely determine the formation of a bow shock.

\begin{table*}[!htb]
\caption{Parameters of [W60]~B90 compared to the three known RSGs with a bow shock}\label{tab:comparative}
\small
\begin{center}
\renewcommand{\arraystretch}{1.4}
\begin{tabular*}{0.95\textwidth}{@{\extracolsep{\fill}}cccccccccc}
\hline \hline
%\toprule
   Name & Sp. type & $T_{\rm{eff,TiO}}$ & $\log(L/\lsun)$ &  Radius & $\dot{M}$ & $P$ & LSP & v$_{\rm pec}$ & Reference \\
   & & (K) & (dex) & (\rsun) & ($10^{-6}~$\msun~yr$^{-1}$) & (d) & (d) & (km s$^{-1}$) & \\
\hline 
  [W60] B90 & M3 I& $3550\pm40$ & $5.32 \pm 0.01$ & $1210$ & $4.4\substack{+5.1 \\ -1.7}$ & 1006 & 4900 & 16-25 & This work, (1) \\  
  Betelgeuse & M2 I & $3650 \pm 25$ & $5.10 \pm 0.22$ & $750-1000$ & $1-4$  & 388 & 2050 & $30$ & (2-5) \\ 
  IRC-10414 & M7 I & $3300$ & $5.2$ & $1200$ & <10 & 768 & - & 70 & (6) \\ 
  $\mu$~Cep & M2 Ia & $3750 \pm 20$ & $5.45 \pm 0.40$ & $1259$ &  7.6$^{a}$ & 860 & 4400 & 80 & (7-9) \\ 
%\bottomrule
\hline
\end{tabular*}
\tablefoot{$^{(a)}$Converted from \cite{Shenoy2016} assuming a gas-to-dust ratio of 200. \\
(1) \cite{Antoniadis2024}, (2) \cite{Levesque2005}, (3) \cite{Joyce2020}, (4) \cite{Kravchenko2021},  (5) 
\cite{LeBertre2012}, (6)~\cite{Gvaramadze2014}, (7) \cite{Josselin2007},  (8) \cite{Shenoy2016} (9) \cite{Tetzlaff2011}}
%periods from Chatys2020
\end{center}
\end{table*}

\subsection{Evolutionary status}\label{sec:dis_HRD}

\begin{figure}
        \centering
   	\resizebox{1.0\hsize}{!}{\includegraphics{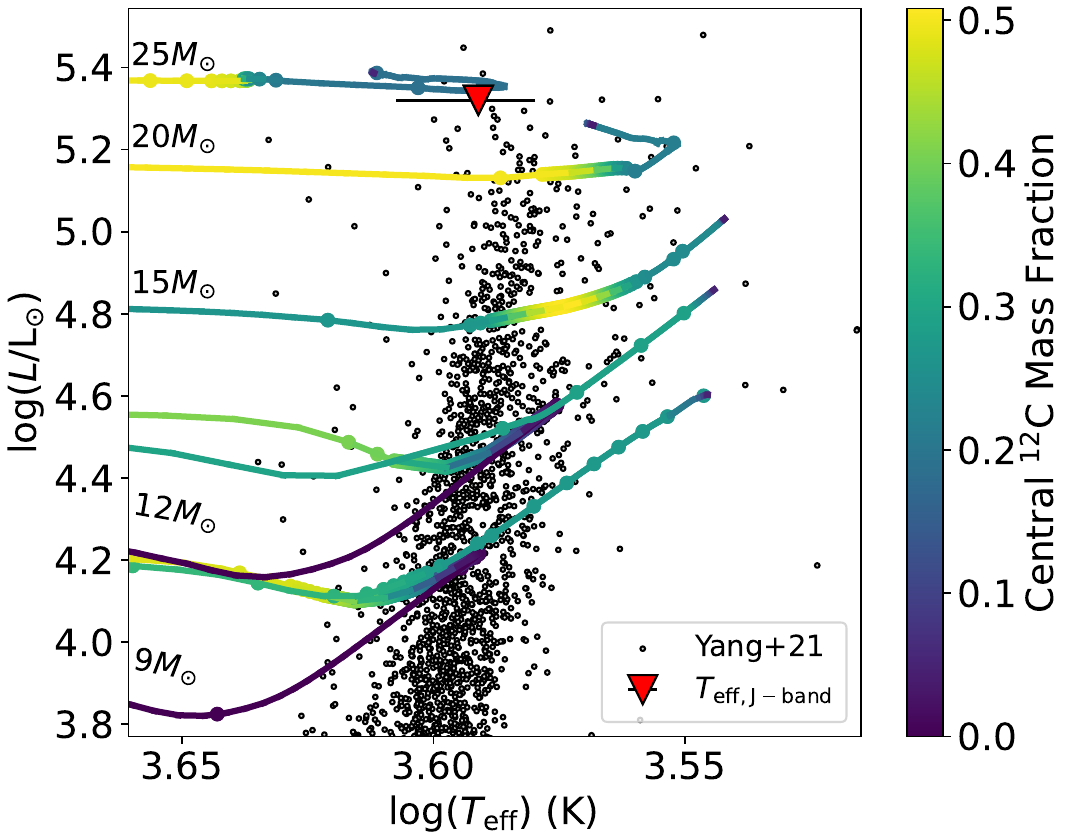}}
  	\caption{Location of [W60]~B90 (red triangle) in the Hertzsprung-Russell diagram of the RSG population in the LMC  \citep[black dots,][]{Yang2021}. The color map represents the central $^{12}$C mass fraction on the MIST evolutionary track, while the nodes indicate a step of 10$^4$ yr. }\label{fig:fig_HRD}
\end{figure}

We explore the location of [W60] B90 in the Hertzsprung-Russell diagram using the derived $T_{\rm{eff,J}}=3900\substack{+150 \\ -100}$~K (see Table~\ref{tab:MARCS_params}) and $\log(L/$\lsun$)=5.32\pm0.01$~dex \citep{Antoniadis2024}. Similar to \citet{deWit2023}, we compare its location to the LMC catalog of RSGs from \citet{Yang2021}. We also used the MIST models \citep{Dotter2016, Choi2016} of rotating single stars ($v= 0.4 v_{\rm crit}$) in the range of 8 to 25 \msun. The position of [W60]~B90 consistently matches the initial mass $M_{\rm ini}=25$~\msun~track, being more massive than Betelgeuse \citep[$M_{\rm ini}=18-21$;][]{Joyce2020}, with current carbon burning in the core and within the last two nodes of the evolutionary track. Therefore, following the expected evolution for a RSG, [W60]~B90 should explode as a Type II SN within the next 10$^{4}$~yr \citep{Smartt2009_evidence_typeII}. However, the observational lack of massive RSGs exploding as SNe \citep[the so-called `RSG problem';][]{Smartt2009, Smartt2015} suggests that either they end their lives in warmer states by stripping part of their envelopes or they collapse into black holes without exploding. Although determining the future evolution of [W60]~B90 is beyond the scope of this paper, monitoring this evolved massive RSG could shed light on the "RSG problem" and indicate whether episodic mass-loss influences the fate of RSGs.

%------------------------------------------------------------------
%------------------------------------------------------------------
%------------------------------------------------------------------
\section{Summary and conclusions}\label{sec:conclusions}

We present a detailed study of the very luminous RSG [W60]~B90 ($\log(L/\lsun)=5.32$~dex), motivated by the discovery of a bar-like structure at 1 pc, which is reminiscent of the bar around Betelgeuse. We found [W60] B90 to be a walkaway star, with a supersonic peculiar velocity between $16-25$~($\pm11$)~km~s$^{-1}$ in the direction of the bar. We also obtained optical long-slit spectroscopy of the circumstellar environment of [W60]~B90 to search for evidence of the hypothesized bow shock. We used the criterion [S~\textsc{ii}]/H$\alpha \!> \!0.4$ to reveal the shocked origin of the nebular emission in the southern part of the bar, which confirms a causal connection with the RSG, and between the bar and the star, where the bow shock is expected. Therefore, [W60]~B90 is the first extragalactic RSG with a suspected bow shock.   

We compiled archival photometry to construct an optical light curve spanning more than 30~yr, reporting three dimming events in the optical with a recurrence of $\sim$11.8~yr and $\Delta V\sim$1~mag. We note a similar recovery timescale of $\sim$400~d for each dimming event in [W60]~B90, in contrast with the Great Dimming of Betelgeuse, which lasted $\sim$200~d. We attribute the delay in the recovery to the size of the atmosphere, as [W60]~B90 is more extended than Betelgeuse and the adjustment within the atmosphere needs additional time to manifest. We support this argument by reporting similarities in the timescale between the dimmings of [W60]~B90 with those of $\mu$~Cep and the hypergiant RW Cep \citep{Anugu2023}, which are comparable in size to our RSG. We also assemble a 10~yr MIR light curve reporting a general amplitude $\Delta \rm W1= 0.51$~mag and $\Delta \rm W2= 0.37$~mag, a variation $\Delta \rm W1= 0.42$~mag and $\Delta \rm W2= 0.25$~mag during the last dimming event, and a long-term variability correlation with the optical. 

Optical multi-epoch spectroscopy during the recovery of the last dimming event revealed different atmospheric properties for each epoch and spectral variability (from M3~I to M4~I), highlighting the importance of light curves in assessing the current state of variable RSGs. We note a correlation correlation between the $T_{\rm{eff,TiO}}$ and the brightness of the star, which might be connected to convection \citep{Kravchenko2019,Kravchenko2021}. We detect an enhancement of $A_V$ after the dimming, which suggests an addition of dust in the line of sight, as a consequence of a mass ejection during the minimum. In addition, a single model cannot reproduce the complex atmosphere of the star during the closest epoch to the minimum. A composite model of cool and hot components considerably improves the description of the spectral features observed. We conclude that [W60]~B90 suffered a mass ejection similar to that reported during the Great Dimming of Betelgeuse \citep{Montarges2021, Dupree2022} and the dimming of RW Cep \citep{Anugu2023}. Furthermore, we find solar-like metallicity $Z=0.0\substack{+0.2 \\ -0.1}$~dex from the atomic lines in the $J$-band, which might indicate the evolved state of [W60]~B90 and prior mixing in the stellar interior entailing an overabundance of metals on the surface. We also report incompatible differences of $\Delta T_{\rm{eff}}>300$~K between the diagnostic in the $J$-band and the optical TiO bands. New models are urgently needed as the current ones are inconsistent depending on the spectral range observed. Further studies are needed in order to construct the new generation of RSG models, allowing further constraint of the basic properties of RSGs such as NLTE assumptions, the wind of the star, convection, and magnetic fields.

The detection of shocked material in the CSM, the high mass-loss rate ($\dot{M}=4.4\substack{+5.1 \\ -1.7}\times10^{-6}~$\msun~yr$^{-1}$), the high variability reported in the optical and the MIR, and the changes in the extinction after the minimum in 2022 suggest that [W60]~B90 is in an unstable evolutionary state and is undergoing episodes of mass loss. Its location in the Hertzsprung-Russell diagram is a good match with the $M_{\rm ini}=25$~ \msun~MIST evolutionary models within the last $10^4$~yr before the end of its life. This work reveals [W60]~B90 to be a perfect laboratory with which to study episodic mass loss in evolved RSGs at low-Z environments. Despite our detailed analysis, further investigation is needed to shed more light on the system. Observations with high spatial resolution, for example with ALMA or VLTI, are needed to resolve the CSM structures and visually identify the speculated bow shock. Additionally, the coronagraph mounted on the \textit{James Webb} Space Telescope would allow us to resolve the closest environment revealing the distribution of the warm dust formed by prior mass ejections. We propose to extend a similar analysis to other very luminous RSGs \citep[e.g., WOH~G64 and RSGs in low-Z galaxies from the ASSESS project;][]{Levesque2009, deWit2024} in order to understand their properties and verify the similarities with respect to [W60]~B90. Constraining the behavior of the most luminous RSGs is crucial for understanding their evolution, as well as the "RSG problem", and the origin of the observed upper $L$ limit of RSGs. 

%--------------------------------------------------------------------

\begin{acknowledgements}
GMS, SdW, AZB, KA, EC, and MK acknowledge funding support from the European Research Council (ERC) under the European Union’s Horizon 2020 research and innovation program (Grant Agreement No. 772086). The work of KB is supported by NOIRLab, which is managed by the Association of Universities for Research in Astronomy (AURA) under a cooperative agreement with the U.S. National Science Foundation. We acknowledge useful discussions with Grigoris Maravelias, Manos Zapartas, Despina Hatzidimitriou, Stavros Akras, Maria Kopsacheili, and Maria del Mar Rubio. We thank the referee, Roberta Humphreys, for all the helpful suggestions. This paper includes data gathered with the 6.5m Magellan Telescopes located at Las Campanas Observatory, Chile. We acknowledge Marcelo Mora and Nidia Morrell for the data acquisition of the Epoch4 and Neb6 spectra obtained with the Baade 6.5m Magellan Telescope. This research is based on observations made with the NASA/ESA Hubble Space Telescope obtained from the Space Telescope Science Institute, which is operated by the Association of Universities for Research in Astronomy, Inc., under NASA contract NAS 5–26555. These observations are associated with program 10583. This work is based in part on observations made with the \textit{Spitzer} Space Telescope, which is operated by the Jet Propulsion Laboratory, California Institute of Technology under a contract with NASA. This work has made use of data from the European Space Agency (ESA) mission \textit{Gaia}~(\url{https://www.cosmos.esa.int/gaia}), processed by the \textit{Gaia} Data Processing and Analysis Consortium (DPAC, \url{https://www.cosmos.esa.int/web/gaia/dpac/consortium}). Funding for the DPAC has been provided by national institutions, in particular, the institutions participating in the \textit{Gaia} Multilateral Agreement. This publication makes use of data products from the Near-Earth Object Wide-field Infrared Survey Explorer (NEOWISE), which is a joint project of the Jet Propulsion Laboratory/California Institute of Technology and the University of Arizona. NEOWISE is funded by the National Aeronautics and Space Administration. This work has made use of data from the Asteroid Terrestrial-impact Last Alert System (ATLAS) project. The Asteroid Terrestrial-impact Last Alert System (ATLAS) project is primarily funded to search for near-earth asteroids through NASA grants NN12AR55G, 80NSSC18K0284, and 80NSSC18K1575; byproducts of the NEO search include images and catalogs from the survey area. This work was partially funded by Kepler/K2 grant J1944/80NSSC19K0112 and HST GO-15889, and STFC grants ST/T000198/1 and ST/S006109/1. The ATLAS science products have been made possible through the contributions of the University of Hawaii Institute for Astronomy, the Queen’s University Belfast, the Space Telescope Science Institute, the South African Astronomical Observatory, and The Millennium Institute of Astrophysics (MAS), Chile. We acknowledge with thanks the variable star observations from the AAVSO International Database contributed by observers worldwide and used in this research. This research made use of Astropy\footnote{\url{http://www.astropy.org}}, a community-developed core Python package for Astronomy \citep{astropy2013, astropy2018}.
\end{acknowledgements}

\bibliographystyle{aa.bst} % style aa.bst
\bibliography{w60b90.bib} % your references Yourfile.bib

%%%%%%%%%%%%%%%%%%
%%% APPENDICES %%%
%%%%%%%%%%%%%%%%%%
\clearpage

\appendix
\onecolumn

\clearpage
\section{Local proper motion values} \label{sec:apendix_PM_values}

Table~A.1 can be found as additional material in: \url{https://zenodo.org/records/13304127}

\section{Apertures and fluxes measured} \label{sec:apendix_aper_flux}
Table~B.1 and B.2 can be found as additional material in: \url{https://zenodo.org/records/13304127}

\begin{figure*}[h]
    \centering
    \includegraphics[width=0.4\textwidth]{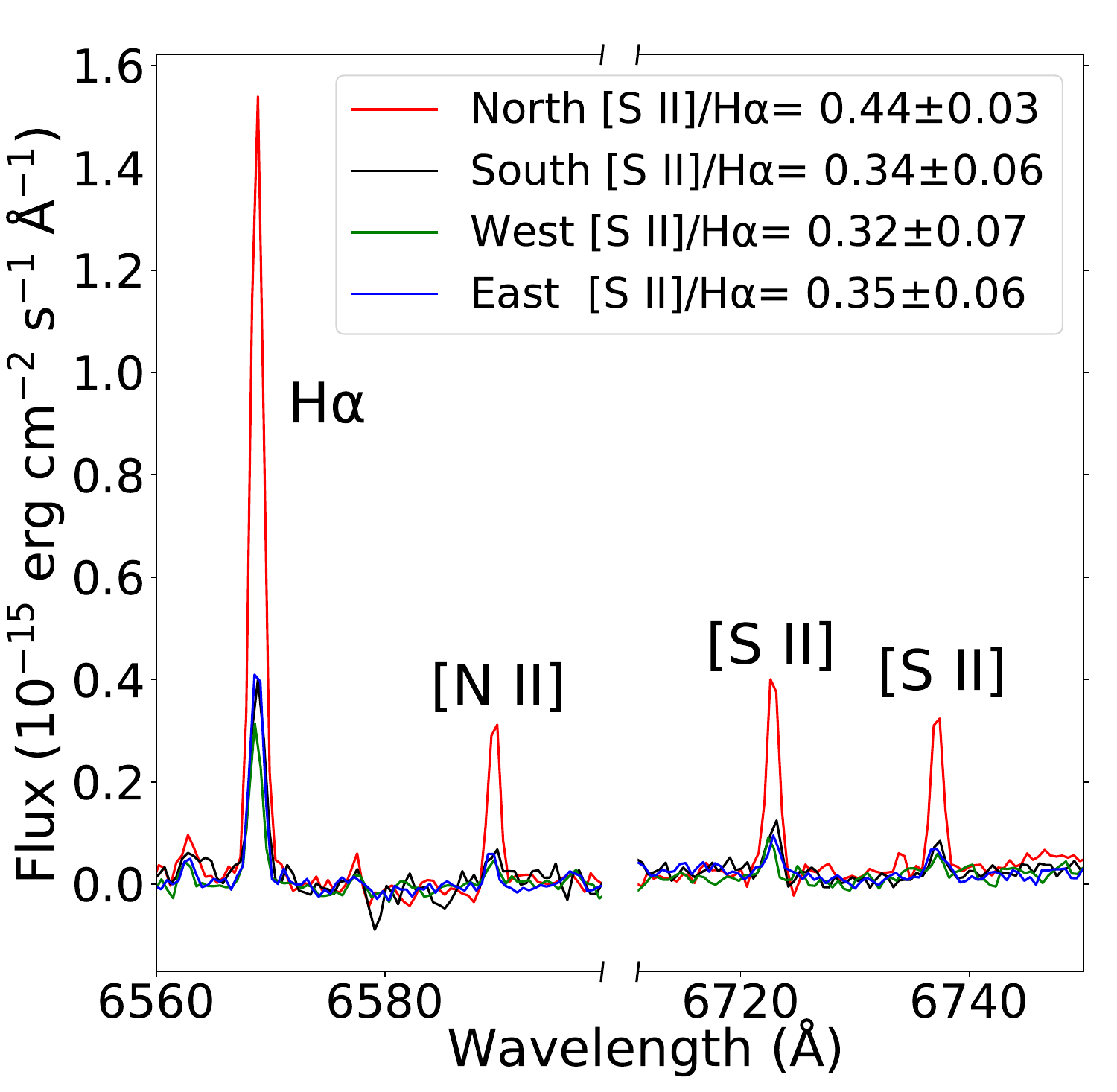}
    \hspace{1cm}
    \includegraphics[width=0.4\textwidth]{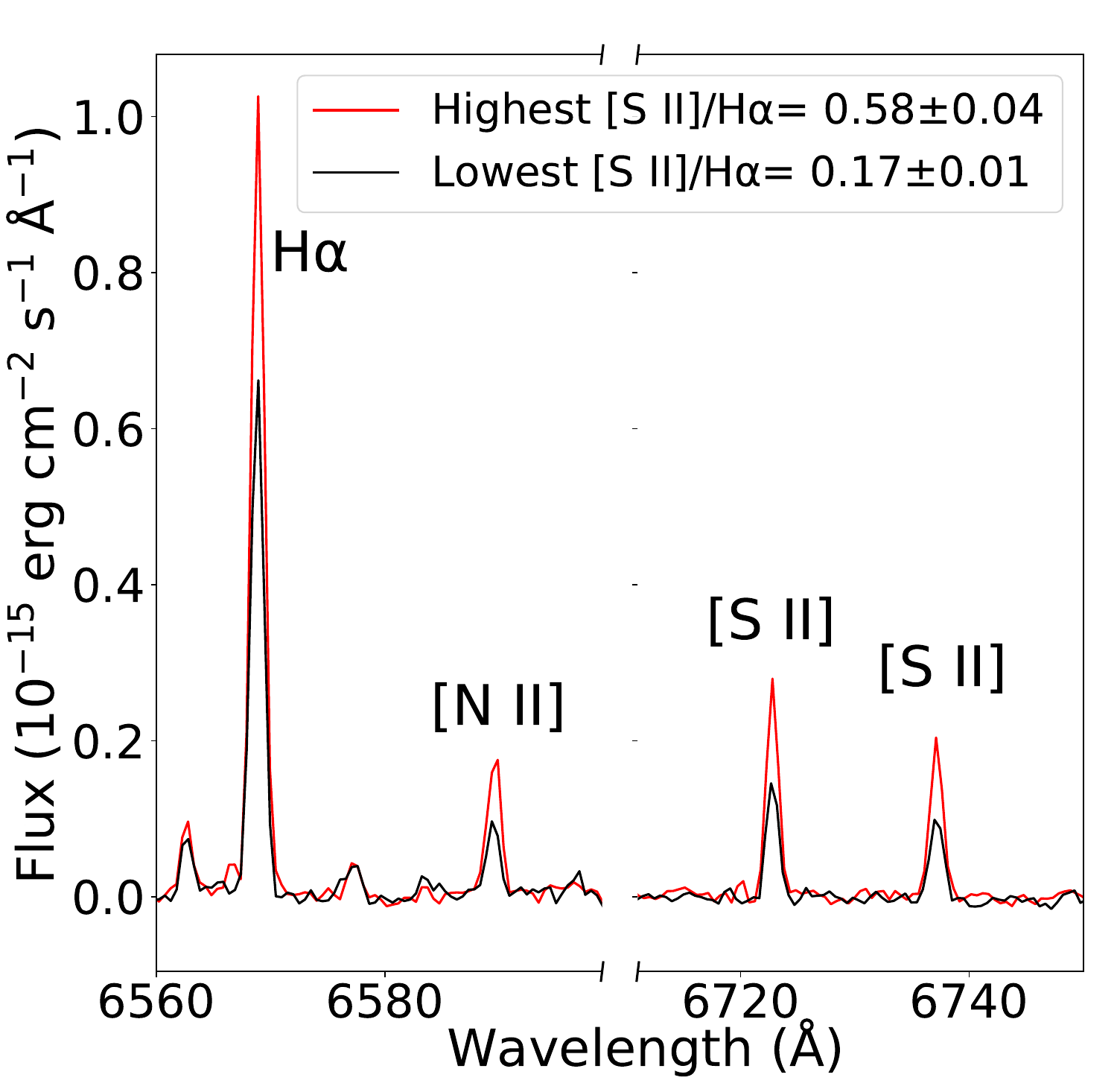}
    \caption{Comparison of the nebular lines in apertures located at 2.1$\arcsec$ in the North, South, East, and West of the star (\textit{left}), and between the apertures with the highest and the lowest [S~\textsc{ii}]/H$\alpha$ ratio (\textit{right}).}\label{fig:examples_SIIvsHa}
\end{figure*}

\clearpage

\section{Spectroscopic analysis of the B-star near [W60]~B90} \label{sec:apendix_Bstar}
\begin{figure}[h]
        \centering
   	\resizebox{0.4\hsize}{!}{\includegraphics{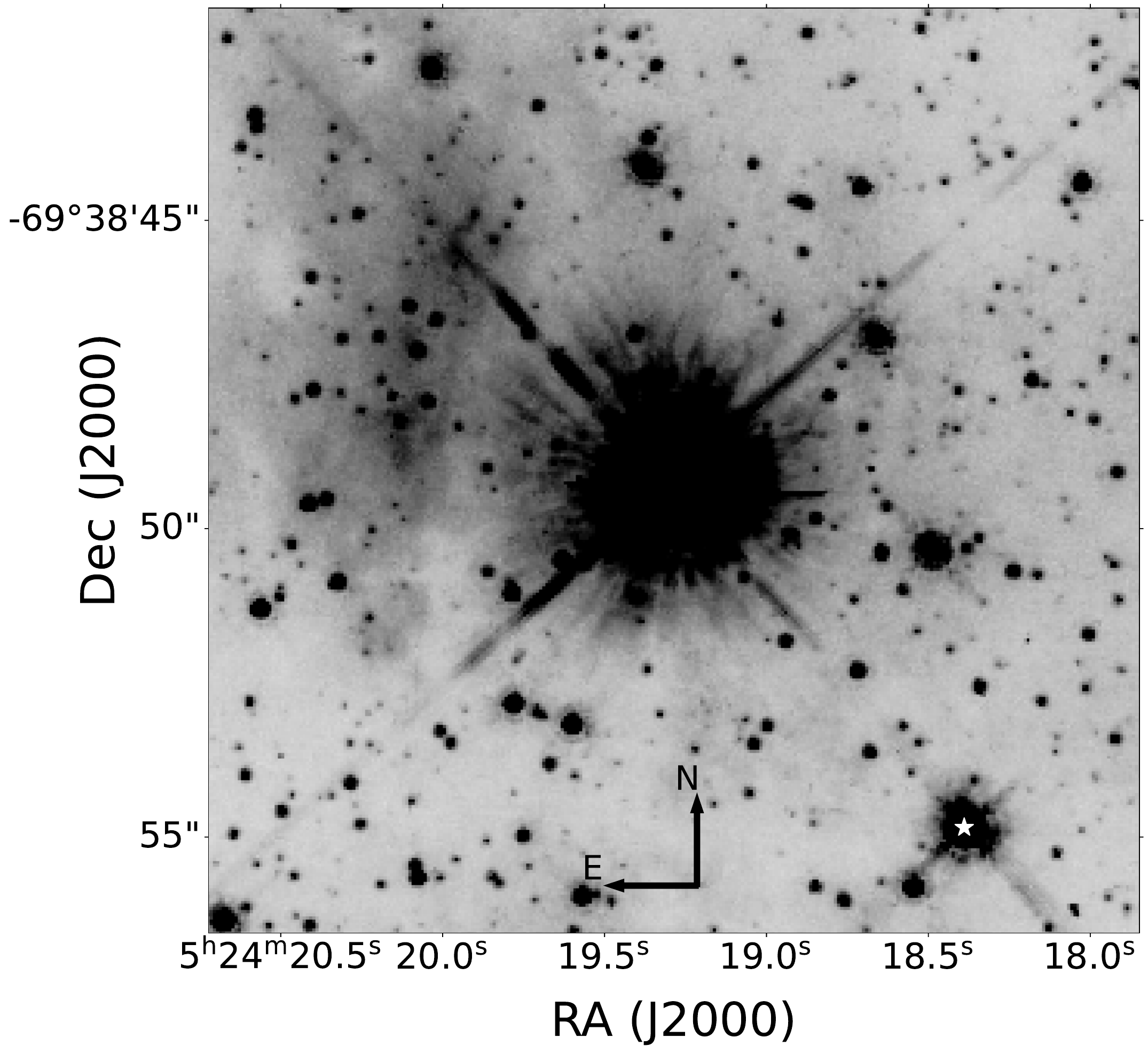}}
  	\caption{\textit{HST} F675W image showing the location of [W60]~B90 and the B-star, marked with a star symbol.}\label{fig:Bstar_loc}
\end{figure}

Apart from the long-slit spectroscopy observations presented in Table~\ref{tab:slit_obs}, we placed the slit in another position that was finally excluded from the analysis of [W60]~B90 and the shocked material. The slit was centered at the coordinates RA=05:24:18.38 Dec= $-$69:38:54.9 on the star with \textit{Gaia}~DR3 ID 4657970853790677248 (Fig.~\ref{fig:Bstar_loc}). The data was obtained on the same night as Epoch4 and Neb6, under the same technical settings, and reduced following the procedure described in Sect~\ref{sec:longslit_obs}. 

\begin{figure}[h]
        \centering
   	\resizebox{1.\hsize}{!}{\includegraphics{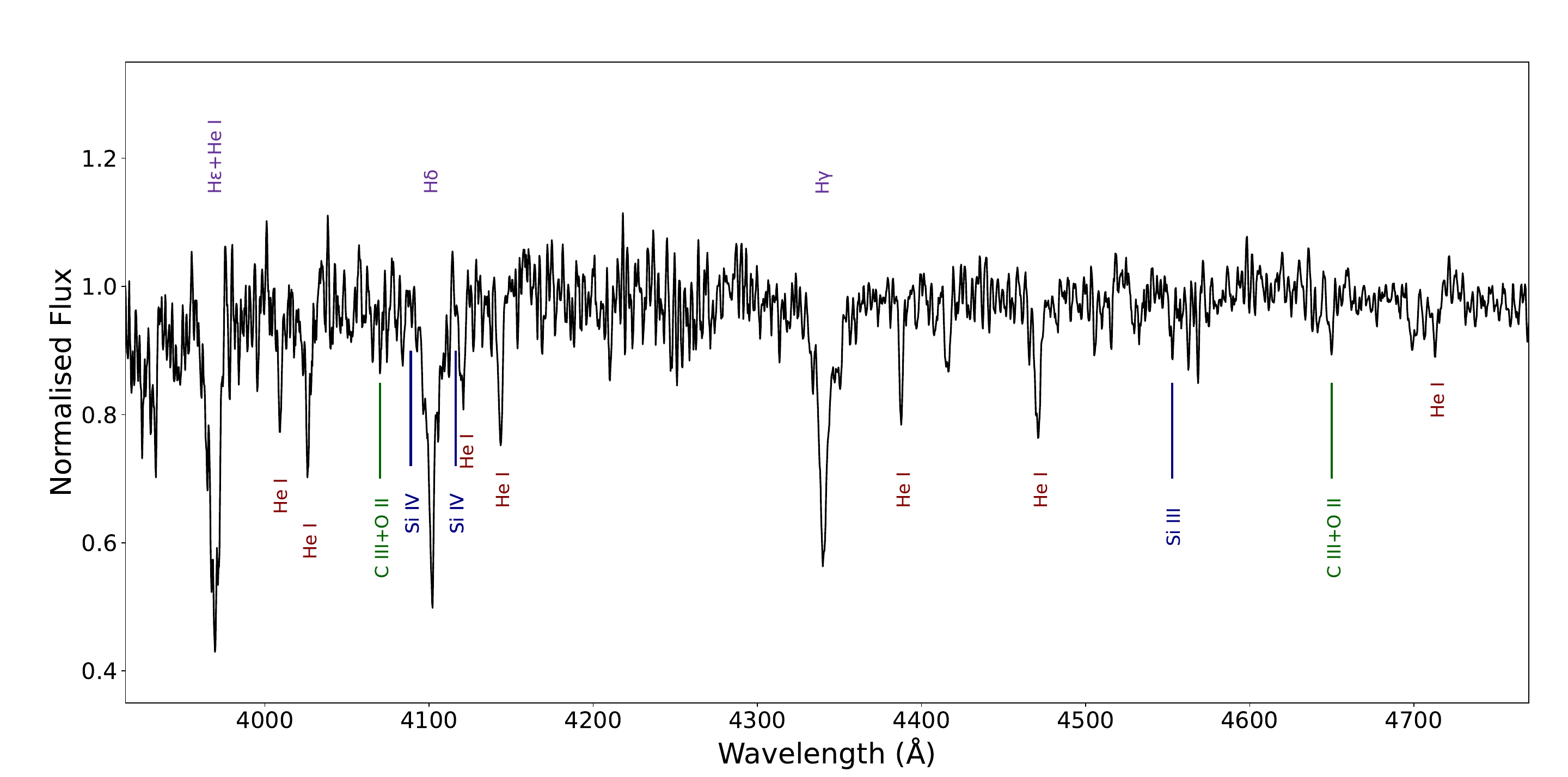}}
  	\caption{Spectral region showing the lines used to determine the B1V spectral type of the star.}\label{fig:Bstar_spec}
\end{figure}

We analyze the spectral region $3900-4800$\r{A} (see Fig.~\ref{fig:Bstar_spec}) using the criteria of \citet{Walborn1990} to determine the spectral classification of the star. The absence of He II lines and the presence of He I lines indicate a B spectral type. More specifically, the absence of He II $\lambda$4686 suggests a type later than B0.7, while the absence of Si II lines $\lambda\lambda4128-4130$ supports a type not later than B2. The weak presence of C III$+$O II blends $\lambda\lambda4070$ and $4650$ suggests a B1 spectral type. The main criterion for the luminosity classification in B1 stars is the weakness of Si III $\lambda4552$ compared to He I $\lambda4387$, which suggests a V luminosity class. This classification is also supported by the low intensity of the Si IV $\lambda4089$ line compared to He I lines $\lambda\lambda4026$ and $4121$, as well as the low ratio of Si IV $\lambda4116$/He I $\lambda4121$. We therefore report a B1V spectral classification for this star. Furthermore, the feature in the right wing of H$\gamma$ and the RV inconsistency in the He I $\lambda6678$ line hint at a low luminosity companion, but further investigation is needed to confirm it. Finally, we find a $P_{LMC}$=0.99 \citep[see Sect.~\ref{sec:proper_motion};][]{Jimenez2023} and a RV of $266\pm7$ km~s$^{-1}$ from the Balmer series, therefore confirming its membership to the LMC.

\clearpage

\section{Disentangling the origin of the nebular emission} \label{sec:appendix_ratios}
\begin{figure*}[ht]
        \centering
        \resizebox{0.87\hsize}{!}{\includegraphics{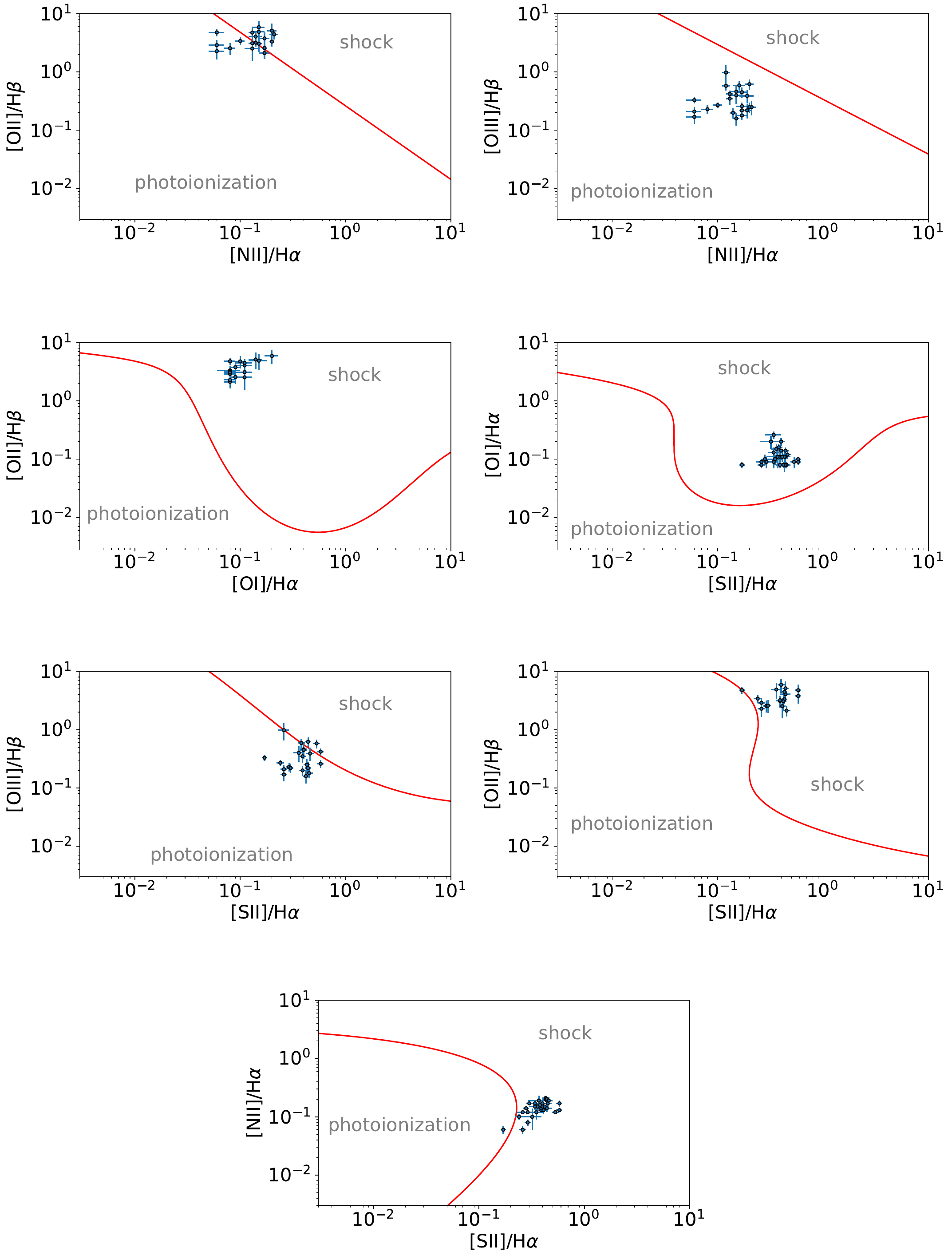}}
  	\caption{Diagnostic plots comparing the line ratio measurements (blue dots) in the CSM around [W60]~B90 with the theoretical predictions of shocked vs. photoionized emission \citep[red line,][]{Kopsacheili2020}.}\label{fig:Kopsacheili_ratios}
\end{figure*}

\clearpage

\end{document}